\documentclass{IEEEtran2}
\ifCLASSINFOpdf
   \usepackage[pdftex]{graphicx}
\else
   \usepackage[dvips]{graphicx}
\fi
\usepackage[cmex10]{amsmath}
\usepackage[utf8]{inputenc}
\usepackage[english]{babel}
\usepackage{multirow}
\usepackage{graphicx}
\graphicspath{ {./images/} }
\usepackage{booktabs}
\usepackage{graphicx}
\usepackage{tabularx}
\usepackage{etoolbox}
\usepackage{nomencl}
\usepackage[linesnumbered,ruled,vlined]{algorithm2e}

\usepackage{lipsum} 

\usepackage{tikz}
\usepackage{array}
\makenomenclature
\usepackage{amsfonts}
\usepackage{adjustbox}
\usepackage{tabularray}
\usepackage{breqn}
\usepackage{color}
\usepackage{makecell}
\usepackage{amsmath}
\usepackage{nicematrix}
\usepackage{cite}
\usepackage{textcomp}
\usepackage{stfloats}
\usepackage{url}
\usepackage{verbatim}
\usepackage{cite}
\usepackage{relsize}
\usepackage[caption=false,font=normalsize,labelfont=sf,textfont=sf]{subfig}
\SetKwInput{KwInput}{Input}                
\SetKwInput{KwOutput}{Output}   
\usepackage{hyphenat}
\usepackage{comment}
\usepackage{multirow}
\usepackage{siunitx} 
\usepackage{booktabs} 
\usepackage{algpseudocode}
\usepackage{rotating}
\usepackage{threeparttable}
\usepackage{comment}
\newif\ifarxiv
\arxivfalse
\SetKwComment{Comment}{/* }{ */}

\renewcommand{\baselinestretch}{0.9825}
\usetikzlibrary{decorations.pathreplacing,calc}
\newcommand{\tikzmark}[1]{\tikz[overlay,remember picture] \node (#1) {};}

\newcommand*{\AddNote}[4]{%
    \begin{tikzpicture}[overlay, remember picture]
        \draw [decoration={brace,amplitude=0.5em},decorate,thick,black]
            ($(#3)!(#1.north)!($(#3)-(0,1)$)$) --  
            ($(#3)!(#2.south)!($(#3)-(0,1)$)$)
                node [align=center, text width=2.5cm, pos=0.5, anchor=west] {#4};
    \end{tikzpicture}
}%

\begin{document}

\title{AC Power Flow Feasibility Restoration via a\\ State Estimation-Based Post-Processing Algorithm}

\vspace{-1em}
\author{Babak Taheri and Daniel K. Molzahn
         }

\maketitle

\begin{abstract}
This paper presents an algorithm for restoring AC power flow feasibility from solutions to simplified optimal power flow (OPF) problems, including convex relaxations, power flow approximations, and machine learning (ML) models. The proposed algorithm employs a state estimation-based post-processing technique in which voltage phasors, power injections, and line flows from solutions to relaxed, approximated, or ML-based OPF problems are treated similarly to noisy measurements in a state estimation algorithm. The algorithm leverages information from various quantities to obtain feasible voltage phasors and power injections that satisfy the AC power flow equations. Weight and bias parameters are computed offline using an adaptive stochastic gradient descent method. By automatically learning the trustworthiness of various outputs from simplified OPF problems, these parameters inform the online computations of the state estimation-based algorithm to both recover feasible solutions and characterize the performance of power flow approximations, relaxations, and ML models. Furthermore, the proposed algorithm can simultaneously utilize combined solutions from different relaxations, approximations, and ML models to enhance performance. Case studies demonstrate the effectiveness and scalability of the proposed algorithm, with solutions that are both AC power flow feasible and much closer to the true AC OPF solutions than alternative methods, often by several orders of magnitude in the squared two-norm loss function. 

\end{abstract}


\thanksto{\noindent 
Submitted to the 23rd Power Systems Computation Conference (PSCC 2024). 
The authors are with the School of Electrical and Computer Engineering, Georgia Institute of Technology, Atlanta, GA, USA. This research was supported by NSF award \#2145564. Provision of data for the machine learning case study results was supported by NSF award \#2112533.}

\vspace{-1.5em}

\section{Introduction}
\label{sec:Introduction}
\IEEEPARstart{O}{ptimal} power flow (OPF) problems seek steady-state operating points for a power system that minimize a specified objective, such as generation costs, losses, voltage deviations, etc., while satisfying equalities that model power flows and inequalities that impose engineering limits. 
The OPF problem forms the basis for many power system applications. Algorithmic improvements have the potential to save billions of dollars annually in the U.S. alone \cite{cain2012history}.
%
%
The AC power flow equations accurately model how a power system behaves during steady-state conditions by relating the complex power injections and line flows to the voltage phasors. Optimal power flow problems that utilize an AC power flow model are not easily solvable and are considered computationally challenging (NP-hard)~\cite{bienstock2019strong}. 

Since first being formulated by Carpentier in 1962 \cite{carpentier1962contribution}, OPF solution methods have been extensively researched~\cite{Momoh1999-1, Momoh1999-2}. With formulations based on the Karush-Kuhn-Tucker (KKT) conditions, many solution methods only seek a local optimum due to the nonconvex nature of the OPF problem. This nonconvexity is due to the nonlinearity of the relationships between active and reactive power injections and voltage magnitudes and angles~\cite{hiskens2001}. Challenges from power flow nonconvexities are further compounded when solving OPF problems that consider, e.g., discreteness and uncertainty~\cite{barrows2014,roald2022review}.
To overcome these challenges, it is common to simplify OPF problems to convex formulations via relaxation and approximation techniques, often resulting in semidefinite programs (SDP)~\cite{lavaei2011zero}, second-order cone programs (SOCP)~\cite{jabr2006radial,coffrin2015qc}, and linear programs~\cite{coffrin2014linear}; see~\cite{molzahn2019} for a survey. A wide range of emerging machine learning (ML) models are also under intense study~\cite{klamkin2022active, pan2022deepopf, chatzos2022, Zamzam2020, kody_chevalier2022}; see~\cite{duchesne2020} for a survey. In this paper, we collectively refer to relaxed, approximated, or ML-based OPF formulations as \emph{simplified} OPF problems.

Relaxed OPF problems bound the optimal objective value, can certify infeasibility, and, when the relaxation is tight, provide globally optimal decision variables~\cite{molzahn2019}. With potential advantages in computational tractability and accuracy when deployed appropriately, power flow approximations are also used in a wide range of operational and design tasks~\cite{molzahn2019}. Whether developed via relaxations or approximations, the convexity of these formulations is valuable in many applications, enabling, for instance, convergence guarantees for distributed optimization algorithms~\cite{molzahn2017survey} and tractability for robust and chance-constrained formulations that consider uncertainties~\cite{molzahn2018towards, Venzke2018}. Similarly, machine learning models hold substantial promise in certain applications, such as quickly characterizing many power injection scenarios with fluctuating load demands and renewable generator outputs~\cite{duchesne2020}.

The computational advantages of simplifying OPF problems via relaxation, approximation, and machine learning techniques come from replacing the nonconvex AC power flow equations with some other model. As a result, all of these simplified OPF problems may suffer from a key deficiency in \emph{accuracy} that is the motivation for this paper. Specifically, the outputs of a simplified OPF problem may not satisfy the AC power flow equations, meaning that the power injections and line flows may be inconsistent with the voltage phasors~\cite{roald2022review,molzahn2019,Venzke2018,bestuzheva2020}. This is problematic since many practical applications for OPF problems require solutions that satisfy the power flow equations to high accuracy.\footnote{Prior research has identified sufficient conditions which ensure the tightness of certain relaxed OPF problems, but the assumptions underlying these conditions make them inapplicable for many practical situations~\cite{molzahn2019,Low2014_exactness}.}

Due to these inaccuracies, there is a need to develop restoration methods that obtain voltage phasors and power injections conforming to the AC power flow equations from the outputs of relaxed, approximated, and ML-based OPF models. There are three main types of methods in the literature that address this. The first type adds penalty terms to the objective function of a relaxed OPF problem. Appropriate choices for these penalty terms can result in the relaxation being tight for the penalized problem, yielding feasible and near-optimal solutions for some OPF problems; see, e.g.,~\cite{madani2014convex}. However, determining the appropriate penalty parameters can be challenging and is often done in a trial-and-error fashion, which can be time-consuming~\cite{venzke2020inexact}. The second type updates the power flow relaxations and approximations within an algorithm that tries to find a local optimum; see, e.g.,~\cite{tian2019recover}, in which the OPF problem is formulated  as a difference-of-convex programming problem and iteratively solved by a penalized convex--concave procedure. As another example, the algorithm in~\cite{fang2021ac} utilizes a power flow approximation to generate an initial operating point and then seeks small adjustments to the outputs of a select number of generators to obtain an operating point that satisfies both the equality and inequality constraints of an OPF problem. These methods can find local optima, but they require good starting points and multiple evaluations of the relaxed or approximate problem, which can be computationally expensive. See~\cite[Ch.~6]{molzahn2019} for a survey of these first two types of methods.

This paper is most closely related to a third type of method that is faster and more direct. This third type of method fixes selected values from the solution of the simplified OPF problem to formulate a power flow problem that has the same number of variables and equations. Solving this power flow problem then yields values for the remaining quantities that satisfy the AC power flow equations. There are various ways to formulate this power flow problem. For instance, one method fixes active power injections and voltage magnitudes at non-slack generator buses to create a power flow problem that can be solved using traditional Newton-based methods~\cite{venzke2020inexact}.  One could instead fix the active and reactive power injections at non-slack generator buses and then solve a power flow problem. As another alternative, one could directly substitute the voltage magnitudes and angles from the simplified OPF solution into the power flow equations to obtain consistent values for active and reactive power generation. 
\textcolor{black}{This third type of method for restoring AC power flow feasibility is commonly used in prior literature, usually as part of a larger algorithm~\cite{venzke2020inexact, Li2022, BOBO2021106625, VANIN2020106699, pan2022deepopf, Zamzam2020}.} Note that this third type of method may result in variable values that do not align with the solution of the simplified problem and may also violate the OPF problem's inequality constraints. Nevertheless, quickly obtaining an AC power flow feasible point that is close to the true OPF solution is often of paramount importance.

Our proposed algorithm for restoring power flow solutions is inspired by ideas from state estimation \cite{abur2004} and offers significantly improved accuracy over alternatives from this third type of method. Instead of fixing some subset of values from a simplified OPF solution, we instead formulate an optimization problem that incorporates additional information regarding the voltage phasors, power injections, and line flows from relaxed, approximated, and ML-based solutions. Our algorithm addresses inaccuracies in power flow models similarly to how measurement errors are handled in state estimation. The use of additional information from simplified solutions allows for a more accurate restoration of the true OPF solution while leveraging mathematical machinery from state estimation. Additionally, our algorithm can simultaneously consider outputs from multiple simplified OPF problems to both improve the quality of the restored solution and automatically characterize the accuracy of different power flow simplifications.

Our proposed algorithm also allows for flexibility in selecting weights, which are comparable to the variances of sensor noise levels in state estimation algorithms, and biases. These weights and biases are parameters that are chosen based on inconsistencies in the solutions to the simplified OPF problems. Determining the best values for these weights and biases is challenging as the inconsistencies in the solutions are not known beforehand. Inspired by ML approaches, we solve this issue via offline training of the weights and biases using the solutions to many actual OPF problems and the corresponding outputs of the simplified OPF problems. To do so, we employ an adaptive stochastic gradient descent (ASGD) algorithm to iteratively update the weights and biases based on information from our proposed restoration algorithm. This minimizes a loss function that compares the actual OPF solutions and the restored points. The trained weights and biases are then used during online calculations to recover the solutions. We demonstrate the proposed restoration algorithm using various convex relaxations, an approximation, and an ML-based model. The results show an improvement of several orders of magnitude in accuracy for some instances. 

In summary, this paper presents an algorithm that addresses AC power flow infeasibility in simplified OPF solutions by:
\begin{itemize}

  \item  Proposing an AC feasibility restoration algorithm relevant to multiple types of simplified OPF problems (relaxations, approximations, and ML models).
  \item  Jointly considering outputs from multiple simplified OPF problems to exploit more information from each.
  \item  Developing an adaptive stochastic gradient descent method to determine optimal weight and bias parameters.
  \item  Illustrating the effectiveness and scalability of the proposed algorithm through numerical experiments using various power flow relaxations, an approximation, and an ML model on multiple test cases.
\end{itemize}
We emphasize that our proposed algorithm does not simply apply least-squares methods to solve power flow problems. Rather, by automatically tuning weighting terms, our proposed approach \emph{learns the trustworthiness of the outputs} from a particular power flow approximation, relaxation, or machine learning model for a specific system and operating range of interest. This enables both \emph{several orders of magnitude improvements in accuracy} for the recovered solutions compared to alternate benchmarks and also provides \emph{targeted insights into the performance} of different approximation, relaxation, and machine learning models with respect to various outputs.

The paper is structured as follows. Section~\ref{sec:Preliminaries} reviews the OPF problem and various simplifications. Section~\ref{sec:Operating Point Recovery Algorithm} presents the proposed restoration algorithm. Section~\ref{sec:Results and Discussion} provides numerical results to demonstrate the algorithm's performance. Section~\ref{sec:Conclusion} concludes the paper and discusses potential avenues for future research. Note that a preliminary version of the proposed algorithm was published in~\cite{taheri2023acc}%
\ifarxiv
\else
\textcolor{black}{~and an extended version of this paper with additional derivations is also available~\cite{Taheri2023arxiv}}.
\fi


\section{OPF Formulation and Simplifications}
\label{sec:Preliminaries}

This section overviews the OPF problem and discusses several common relaxations and approximations as well as emerging machine learning models that simplify the OPF problem to improve tractability at the cost of accuracy. For more detail, see~\cite{molzahn2019} for a survey on power flow relaxations and approximations and~\cite{duchesne2020} for a survey on machine learning.

We first establish notation. The sets of buses and lines are represented by $\mathcal{N}$ and $\mathcal{E}$, respectively. Each bus $i\in\mathcal{N}$ has a voltage phasor $V_i$ with phase angle $\theta_{i}$, a complex power demand $S^d_i$, a shunt admittance $Y_i^S$, and a complex power generation $S^g_i$. Buses without generators are modeled as having zero generation limits. Complex power flows into each terminal of each line $(j,k)\in\mathcal{E}$ are denoted as $S_{jk}$ and $S_{kj}$. Each line $(j,k)\in\mathcal{E}$ has admittance parameters $Y_{jk}$ and $Y_{kj}$. The real and imaginary parts of a complex number are denoted as $\Re(\,\cdot\,)$ and $\Im(\,\cdot\,)$, respectively. The complex conjugate of a number is represented by $(\,\cdot\,)^*$ and the transpose of a matrix is represented by $(\,\cdot\,)^T$. Upper and lower bounds are denoted as $(\overline{\,\cdot\,})$ and $(\underline{\,\cdot\,})$, interpreted as separate bounds on real and imaginary parts for complex variables. The OPF problem is:
\begin{subequations}
\label{eq:opf}
\begin{align}
& \label{eq:ac_pf1}\min \sum_{i \in N} c_{2 i}\left(\Re\left(S_{i}^{g}\right)\right)^{2}+c_{1 i} \Re\left(S_{i}^{g}\right)+c_{0 i} \hspace*{-10em} \\
\nonumber & \text{s.t.} ~~~ (\forall i\in\mathcal{N}, ~\forall (j,k)\in\mathcal{E})\\
& \label{eq:ac_pf8} \mathbf{W}_{jk}=V_j^{\vphantom{*}} V^{*}_k, ~ \mathbf{W}_{kj}=V_k^{\vphantom{*}} V^{*}_j, ~ \mathbf{W}_{ii}=V_i^{\vphantom{*}} V^{*}_i \\
& \label{eq:ac_pf9} \left(\underline{V}_{i}\right)^{2} \leq \mathbf{W}_{i i} \leq\left(\overline{V}_{i}\right)^{2} \\
& \label{eq:ac_pf10} \underline{S}^{g}_i \leq S^{g}_i \leq \overline{S}^{g}_i \\
& \label{eq:ac_pf11} \left|S_{jk}\right| \leq \overline{S}_{jk}, ~\left|S_{kj}\right| \leq \overline{S}_{jk}\\
& \label{eq:ac_pf12} S^{g}_i-S^{d}_i - \left(Y_i^S\right)^* \mathbf{W}_{ii}=\!\!\sum_{(i, j) \in \mathcal{E}}\!\! S_{i j} + \!\!\sum_{(k, i) \in \mathcal{E}}\!\! S_{ki}\\
& \label{eq:ac_pf13}   S_{jk}=Y_{jk}^{*} \mathbf{W}_{jj}-Y_{jk}^{*} \mathbf{W}_{jk} \\
& \label{eq:ac_pf14} S_{kj}=Y_{kj}^{*} \mathbf{W}_{kk}-Y_{kj}^{*} \mathbf{W}_{kj}^{*} \\
& \label{eq:ac_pf15} \tan \left(-\overline{\theta}_{jk}\right) \Re\left(\mathbf{W}_{jk}\right) \leq \Im\left(\mathbf{W}_{jk}\right) \leq \tan \left(\overline{\theta}_{jk}\right) \Re\left(\mathbf{W}_{jk}\right).
\end{align}
\end{subequations}
 
The OPF problem~\eqref{eq:opf} minimizes an objective, in this case, the generation cost as shown by~\eqref{eq:ac_pf1}. The objective has quadratic coefficients $c_{2i}$, $c_{1i}$, and $c_{0i}$. The voltage phasor products are collected in a Hermitian matrix~$\mathbf{W}$, as described in~\eqref{eq:ac_pf8}. The OPF problem also imposes voltage magnitude limits~\eqref{eq:ac_pf9}, generator output limits~\eqref{eq:ac_pf10}, apparent power flow limits~\eqref{eq:ac_pf11}, and complex power balance at each bus~\eqref{eq:ac_pf12}. Power flows for each line are defined in~\eqref{eq:ac_pf13} and~\eqref{eq:ac_pf14} and limits on phase angle differences across lines are imposed in~\eqref{eq:ac_pf15}. 

All the nonconvexities in the OPF problem are associated with the products in~\eqref{eq:ac_pf8}, which, in combination with~\eqref{eq:ac_pf12}--\eqref{eq:ac_pf14}, form the power flow equations. Power flow relaxations convexify the power flow equations by replacing~\eqref{eq:ac_pf8} with less stringent conditions. The SDP relaxation requires that $\mathbf{W}$ is positive semidefinite~\cite{lavaei2011zero}, which is implied by~\eqref{eq:ac_pf8}. The SOCP relaxation requires that $\mathbf{W}$ has non-negative principal minors~\cite{jabr2006radial}, which is implied by positive semidefiniteness of $\mathbf{W}$. The QC relaxation strengthens the SOCP relaxation with additional variables and constraints corresponding to phase angle differences that are restricted to convex envelopes~\cite{coffrin2015qc}. 

Power flow approximations replace~\eqref{eq:ac_pf8}, \eqref{eq:ac_pf12}--\eqref{eq:ac_pf14} with alternative (usually linear) expressions relating the line flows and voltage phasors. For instance, the linear-programming approximation (LPAC) linearizes a polar representation of the power flow equations~\cite{coffrin2014lpac}. This is accomplished by linearizing sine functions using a first-order Taylor approximation around zero phase angle differences and replacing cosine functions with lifted variables restricted to a convex polytope. 

Some emerging ML approaches for OPF problems replace the power flow equations~\eqref{eq:ac_pf8}, \eqref{eq:ac_pf12}--\eqref{eq:ac_pf14} with a surrogate model. For instance, the approach in~\cite{kody_chevalier2022} replaces the power flow equations with the piecewise linear function corresponding to a trained neural network with rectified linear unit (ReLU) activation functions to obtain a mixed-integer linear programming formulation. Other ML approaches such as~\cite{klamkin2022active, pan2022deepopf, chatzos2022, Zamzam2020} directly estimate OPF solutions by training neural networks or other ML models to predict values for the optimal voltage phasors, power injections, and/or line flows.

Since these simplified OPF formulations do not enforce the true AC power flow equations~\eqref{eq:ac_pf8}, \eqref{eq:ac_pf12}--\eqref{eq:ac_pf14}, their outputs may have inconsistencies between the power injections, power flows, and voltage phasors. Thus, all would potentially benefit from our proposed restoration algorithm. We next present our proposed restoration algorithm considering a generic simplified OPF problem. In Section~\ref{sec:Results and Discussion}, we illustrate the method's performance using the SDP, SOCP, and QC relaxations, the LPAC approximation, and the ML model from~\cite{klamkin2022active}.


\section{Restoring AC Power Flow Feasibility}
\label{sec:Operating Point Recovery Algorithm}


Solutions to any of the simplified OPF problems discussed in Section~\ref{sec:Preliminaries} may suffer from voltage phasors, power injections, and line flows that do not satisfy the AC power flow equations. To restore operating points that are AC power flow feasible, this section introduces a restoration algorithm inspired by state estimation techniques where the voltage phasors, power injections, and line flows from the simplified solution are analogous to noisy measurements. This algorithm improves on previous methods as it does not fix any variables to specific values, thus enabling the restoration of higher-quality solutions. Note that this algorithm \emph{does not rely on actual measurements} from the physical system. Instead, this algorithm finds the AC power flow feasible voltage phasors that most closely match the voltage phasors, power injections, and line flows resulting from the solution to a simplified OPF problem in a similar manner by which state estimation algorithms resolve inconsistencies among noisy measurements.

\begin{figure}[!t]
    \centering
    \includegraphics[width=8cm]{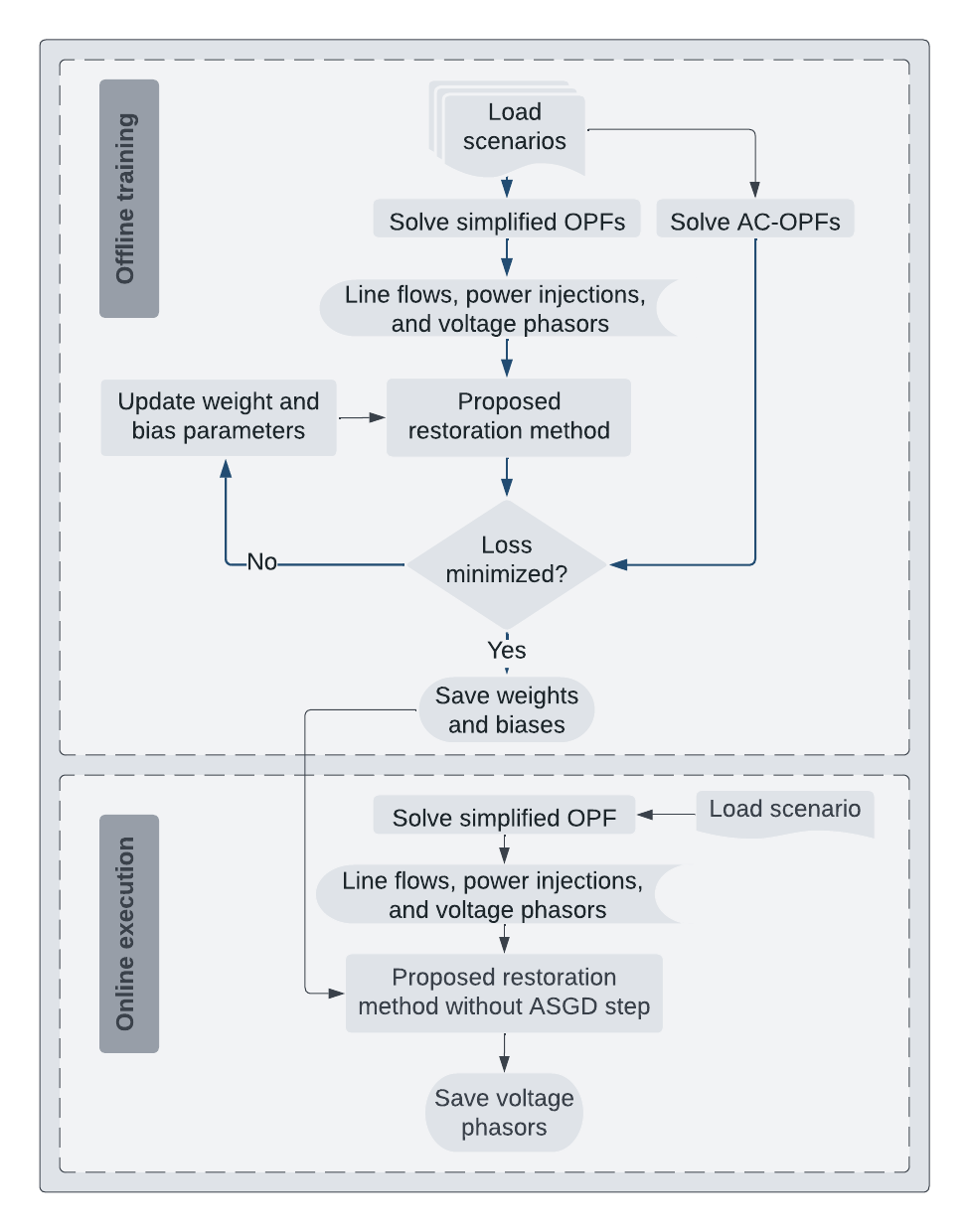}
    \vspace{-1em}
     \caption{Flowchart of the proposed algorithm.}
     \label{fig:flowchart}
    \vspace{-1.5em}
 \end{figure}

Analogous to how state estimation algorithms use variations in the amount of sensor noise to weight measured quantities, the proposed algorithm includes weight and bias parameters  associated with the outputs of each quantity from the simplified OPF solution. However, unlike state estimation algorithms, these weight parameters are not determined by the physical characteristics of a sensor, but rather by the inconsistencies (with regard to the AC power flow equations) among various quantities in the solution to the simplified OPF problem. To determine the optimal values for these weight and bias parameters, we propose an ASGD-based method that is executed offline, with the results used online for restoring AC power flow feasibility. 
Fig.~\ref{fig:flowchart} shows both the algorithm for determining the weights and biases and the solution restoration algorithm. Furthermore, Table \ref{tab:analogy} summarizes the analogy between the proposed algorithm and state estimation.

\subsection{AC Feasibility Restoration Algorithm}

In this section, we introduce our proposed algorithm for restoring AC power flow feasible points from the solutions of simplified OPF problems (convex relaxations, approximations, and ML-based models). This method aims to find the voltage phasors that are close to the true OPF solution's voltage phasors based on the voltage phasors, power injections, and line flows from a simplified OPF solution.

\begin{table}[t]
\caption{Analogy relating the proposed algorithm and state estimation}
\vspace{-0.5em}
\label{tab:analogy}
\centering
\renewcommand{\arraystretch}{1} 
\begin{tabular}{|c|c|} 
 \hline 
 \thead{\textbf{Proposed Algorithm}} & \thead{\textbf{State Estimation}}\\ 
 \hline\hline
 \thead{Solutions from relaxed, \\approximated, or ML-based models} & \thead{Measurements from \\physical sensors} \\ 
 \hline
 \thead{Inconsistencies in relaxed, \\ approximated, or ML-based solutions} & \thead{Noise from physical sensors} \\ 
 \hline
\thead{Weight parameters} & \thead{Variance of the \\  measurement noise} \\
\hline
\thead{Bias parameters} & ---\\
\hline
\end{tabular}
\vspace{-2em}
\end{table}

This section introduces notation based on the typical representation of state estimation algorithms \cite{abur2004} to show how this mathematical machinery is leveraged in our method. We emphasize that we do not use any actual measurements from physical sensors, but rather use the information from the simplified OPF solution. The goal is to find the voltage magnitudes and angles (denoted as $\mathbf{x}$) that are most consistent with the voltage magnitudes, phase angles, power flows, and power injections from the simplified OPF solution, which we gather into a vector $\mathbf{z}$. We denote the number of these quantities, i.e., the length of $\mathbf{z}$, by $m$ and let $n$ be the number of voltage magnitudes plus the number of (non-slack) voltage angles, i.e., the length of~$\mathbf{x}$.

We also define a length-$m$ vector of bias parameters $\mathbf{b}$ for the simplified solution.\footnote{As we will discuss in Section~\ref{subsec:combining}, we can generalize our algorithm to jointly consider solutions from multiple simplified OPF problems. In this case, the elements of $\mathbf{z}$ and $\mathbf{b}$ correspond to quantities from multiple simplified OPF problems stacked together, but the size of $\mathbf{x}$ remains the same. }  
\textcolor{black}{While measurement errors in state estimation are typically only characterized by their variations, the errors in simplified OPF solutions may be \emph{biased}, i.e., consistently overestimate or underestimate the true values of some quantities. We account for this using a bias term $\mathbf{b}$ that represents the systematic errors in the simplified OPF solutions. We also have an error term $\mathbf{e}$ that captures variations that are modeled as being random. By considering both bias and error terms, our model is better equipped to handle both systematic and random deviations from the true values.} We use an AC power flow model denoted as $\mathbf{h}(\mathbf{x})$ to relate $\mathbf{x}$ and $\mathbf{\mathbf{z}}$, parameterized by the bias $\mathbf{b}$:
%
\begin{equation}
z_i+b_i=h_i(\mathbf{x})+e_i, \quad i=1,\ldots, m,
\end{equation}
where $\mathbf{h(x)}= [\mathbf{V(x)}^T~\mathbf{P(x)}^T~\mathbf{Q(x)}^T~\mathbf{P}^f\!\mathbf{(x)}^T~\mathbf{Q}^f\!\mathbf{(x)}^T~\boldsymbol{\theta}(\mathbf{x})^T]^T $  denotes the AC power flow equations relating the vectors of voltage magnitudes $\mathbf{V}$, active and reactive power injections $\mathbf{P}$ and $\mathbf{Q}$, active and reactive line flows $\mathbf{P}^f$ and $\mathbf{Q}^f$, and voltage angles $\boldsymbol{\theta}$ to the vector $\mathbf{x}= [\boldsymbol{\theta}^T~\mathbf{V}^T]^T $. The first and last entries of $\mathbf{h(x)}$, namely, $\mathbf{V(x)}$ and $\boldsymbol{\theta}\mathbf{(x)}$, are obtained via the identity function. The remaining entries of the $\mathbf{h(x)}$ are:
%

%
%
\begin{subequations}\label{eq:h}
\begin{align}
P_i =& \sum_{(i,j) \in \mathcal{E}} P^f_{ij}, \qquad  Q_i = \sum_{(i,j) \in \mathcal{E}} Q^f_{ij},  \\
P^f_{i j} =& V_i^2\left(\Re(Y_{ij})+\Re(Y_{ij}^{sh})\right)- V_i V_j \Re(Y_{ij}) \cos (\theta_{i}-\theta_{j})\nonumber
&\\ 
& \quad  - V_i V_j \Im(Y_{ij}) \sin (\theta_{i}-\theta_{j}), \\
Q^f_{i j} =& -V_i^2\left(\Im(Y_{ij}) +\Im(Y_{ij}^{sh}) \right) - V_i V_j \Re(Y_{ij})  \sin (\theta_{i}-\theta_{j})\nonumber &\\
& \quad + V_i V_j\Im(Y_{ij})  \cos (\theta_{i}-\theta_{j}).
\end{align}
\end{subequations}
\textcolor{black}{Hence, the error $\mathbf{e}$ is the difference between the simplified solution (offset by the bias parameter) and the value corresponding to the restored point $\mathbf{x}$}.
%
As we will discuss below in Section~\ref{subsec:weights}, the bias~$\mathbf{b}$ is computed offline based on the characteristics of many simplified OPF solutions to reflect the systematic offsets of the simplified solutions from the true values. After determining the bias $\mathbf{b}$, the error $\mathbf{e}$ represents the remaining inconsistencies that are not accounted for by the systematic bias.

To address these remaining inconsistencies, our proposed restoration algorithm uses a weighted least squares formulation similar to typical state estimation algorithms. The goal is to choose the voltage magnitudes and angles in $\mathbf{x}$ that minimize a cost function, denoted as $J(\mathbf{x})$, that is the sum of the squared inconsistencies between the simplified OPF solution (offset by~$\mathbf{b}$) and the true OPF solution, i.e., the difference between $\mathbf{h}(\mathbf{x})$ and $\mathbf{z}+\mathbf{b}$. These inconsistencies are represented by the vector $\mathbf{e}$ and are weighted by a specified diagonal matrix $\mathbf{\Sigma}$ with weight parameters associated with the measurements from the simplified OPF solution:
%
\begin{equation}\label{eq:se_setup1}
 \min\, J(\mathbf{x}) = \mathbf{e}^{T} \mathbf{\Sigma} \mathbf{e}.
\end{equation}
In a state estimation application, $\mathbf{\Sigma}$ would be the covariance matrix for the sensor noise. Conversely, we permit $\mathbf{\Sigma}$ to be any diagonal matrix with values computed offline using the algorithm described in Section~\ref{subsec:weights}.


We solve~\eqref{eq:se_setup1} by considering the optimality conditions:
%
\begin{equation}\label{eq:se_g}
    \mathbf{g}(\mathbf{x})=\frac{\partial J(\mathbf{x})}{\partial \mathbf{x}}=-\mathbf{\mathbf{H}}(\mathbf{x})^{T} \mathbf{\Sigma} (\mathbf{\mathbf{z}+\mathbf{b}}-\mathbf{h}(\mathbf{x}))=\mathbf{0},
\end{equation}
where the Jacobian matrix $\mathbf{\mathbf{H}}(\mathbf{x})$ of AC power flow equations $\mathbf{h}(\mathbf{x})$ is $\mathbf{\mathbf{H}}(\mathbf{x})=\frac{\partial \mathbf{h}(\mathbf{x})}{\partial \mathbf{x}}$:
%
%
\begin{equation}
\mathbf{H(x)} =
\begin{bmatrix}
\mathbf{0} & \frac{\partial \mathbf{P}}{\partial \boldsymbol{\theta}} & \frac{\partial \mathbf{Q}}{\partial \boldsymbol{\theta}} & \frac{\partial \mathbf{P^f}}{\partial \boldsymbol{\theta}} & \frac{\partial \mathbf{Q^f}}{\partial \boldsymbol{\theta}} & \mathbf{1} \\
\mathbf{1} & \frac{\partial \mathbf{P}}{\partial \mathbf{V}} & \frac{\partial \mathbf{Q}}{\partial \mathbf{V}} & \frac{\partial \mathbf{P^f}}{\partial \mathbf{V}} & \frac{\partial \mathbf{Q^f}}{\partial \mathbf{V}} & \mathbf{0} \
\end{bmatrix}^T.
\end{equation}

To compute the solution to~\eqref{eq:se_g}, we apply the Newton-Raphson method described in Algorithm~\ref{alg:NR} that solves, at the $k$-th iteration, the following linear system:
%
\begin{equation}
\label{eq:newton0}
\mathbf{x}^{(k+1)}=\mathbf{x}^{(k)}-(\mathbf{G}(\mathbf{x}))^{-1}\mathbf{g}(\mathbf{x}),
\end{equation}
where 
%
\begin{equation}
    \mathbf{G}(\mathbf{x})=\frac{\partial \mathbf{g}(\mathbf{x})}{\partial \mathbf{x}}= \mathbf{H(x)}^T \mathbf{\Sigma} \mathbf{H(x)}
\end{equation}
%
Algorithm~\ref{alg:NR} uses a convergence tolerance of $\epsilon$ and takes a user-specified initialization of $\mathbf{x}^{(0)}$. When available, the voltage magnitudes and angles from the simplified OPF solution often provide reasonable initializations $\mathbf{x}^{(0)}$. Otherwise, a flat start provides an alternative initialization. The output of this algorithm is the restored solution, denoted as $\mathbf{x}_{R}$.

\begin{algorithm}[t]
\caption{Newton-Raphson Algorithm for AC Power Flow Feasibility Restoration}
\label{alg:NR}
\smaller 
\SetKwInOut{Input}{Input}
\SetKwInOut{Output}{Output}
\SetKw{Initialize}{Initialize}
\SetKw{Calculate}{Calculate}
\Input{Simplified OPF solution $\mathbf{z}$, Initialization $\mathbf{x}^{(0)}$, Parameters $\mathbf{\Sigma}$, $\mathbf{b}$, $\epsilon$}
\Output{Restored AC power flow feasible solution $\mathbf{x}_{R}$}
\Initialize{$k \leftarrow 0$}

\While{$||\boldsymbol{\Delta x}^{(k)}|| > \epsilon$}{
$k \leftarrow k + 1$\;
\Calculate{$\mathbf{h}(\mathbf{x}^{(k)})$, $\mathbf{H}(\mathbf{x}^{(k)})$, $\mathbf{G}(\mathbf{x}^{(k)})$, and $\mathbf{g}(\mathbf{x}^{(k)})$ using $\mathbf{\Sigma}$, $\mathbf{z}$, $\mathbf{b}$, and $\mathbf{x}^{(k)}$}\;
$\begin{aligned}
\boldsymbol{\Delta x} ^{(k)}=&\mathbf{G}(\mathbf{x}^{(k)})^{-1} \mathbf{H}(\mathbf{x}^{(k)})^{T} \mathbf{\Sigma}\left(\mathbf{z}+\mathbf{b}-\mathbf{h}(\mathbf{x}^{(k)})\right);
\end{aligned}$

$\mathbf{x}^{(k+1)} \leftarrow \mathbf{x}^{(k)} + \boldsymbol{\Delta x}^{(k)}$;

$\mathbf{x}_{R} = \mathbf{x}^{(k+1)}$ 
}
\end{algorithm}

\subsection{Determining the Weight Parameters}\label{subsec:weights}

The weight parameters $\mathbf{\Sigma}$ play a crucial role in determining the accuracy of the solution obtained from the Algorithm~\ref{alg:NR}. Ideally, larger values of~$\mathbf{\Sigma}_{ii}$ should be chosen for the quantities~$\mathbf{z}_i$ from the simplified OPF solution that more closely represent the solution to the true OPF problem. However, it is not straightforward to predict or estimate the accuracy of a particular $\mathbf{z}_i$ in relation to the true OPF solution. Choosing values for the bias parameters $\mathbf{b}$ poses similar challenges. As shown in Fig.~\ref{fig:flowchart}, we therefore develop an approach inspired by the training of machine learning models to determine these parameters. This approach involves solving a set of randomly generated OPF problems along with the corresponding simplified OPF problems to create a training dataset. As presented in Algorithm~\ref{alg:proposed-method}, we then employ an ASGD method that iteratively solves the proposed restoration algorithm and updates the weight parameters in a way that minimizes the difference between the restored solution and the true OPF solution across the training dataset. 
\textcolor{black}{In this offline training, solutions to each of the true and simplified OPF problems are computed in parallel.}

The ASGD method relies on the sensitivities of the restored point $\mathbf{x}_R$ with respect to the parameters $\mathbf{\Sigma}$, i.e., $\frac{ \partial \mathbf{x}_{R}}{\partial \mathbf{\Sigma}}$:
%
\begin{align}
\label{vector}
\frac{\textrm{vec}(\partial \mathbf{x}_{R})}{\textrm{vec}(\partial \mathbf{\Sigma})} &= \bigg((\mathbf{z}+\mathbf{b}-\mathbf{h(x)}) - \Big(\mathbf{\mathbf{H(x)}}(\mathbf{\mathbf{H(x)}}^T \mathbf{\Sigma} \mathbf{\mathbf{H(x)}} )^{-1} \nonumber \\
&\qquad \quad \times \mathbf{\mathbf{H(x)}}^T \mathbf{\Sigma} (\mathbf{z}+\mathbf{b}-\mathbf{h(x)})\Big)\bigg)  \nonumber \\
&\qquad \otimes \bigg((\mathbf{\mathbf{H(x)}}^T \mathbf{\Sigma} \mathbf{\mathbf{H(x)}})^{-1} \mathbf{H(x)}^T \bigg)^T.
\end{align}
The expression~\eqref{vector} gives the sensitivities of the restored point $\mathbf{x}_R$ with respect to the weight parameters $\mathbf{\Sigma}$, where $\otimes$ denotes the Kronecker product and $\textrm{vec}(\,\cdot\,)$ denotes the vectorization of a matrix.
With length-$m$ vectors $\mathbf{z}$ and $\mathbf{h}(\mathbf{x})$ and an $m\times n$ matrix $\mathbf{H}$, the sensitivities $\frac{ \partial \mathbf{x}_{R}}{\partial \mathbf{\Sigma}}$ are represented by a $n \times m^2$ matrix.
Appendix~\ref{sec:sensitivity_derivation} provides the derivation of~\eqref{vector}.

Note that directly applying~\eqref{vector} can be computationally expensive for large-scale systems since this expression computes sensitivities with respect to all entries of $\mathbf{\Sigma}$. While possibly relevant for variants of the proposed formulation, sensitivities for the off-diagonal terms in $\mathbf{\Sigma}$ are irrelevant in our formulation since these terms are fixed to zero due to the diagonal structure of $\mathbf{\Sigma}$. Exploiting this structure can enable faster computations;
see Appendix~\ref{sec:diagonal} for further details.

\subsection{Determining the Bias Parameters}
Like the weight parameters $\mathbf{\Sigma}$, the bias parameters $\mathbf{b}$ significantly impact the accuracy of Algorithm~\ref{alg:NR}'s outputs. Similar to the weight parameters $\mathbf{\Sigma}$, Algorithm~\ref{alg:proposed-method} also computes the bias parameters $\mathbf{b}$ via an ASGD method. The sensitivities of the restored point $\mathbf{x}_R$ with respect to the bias parameters $\mathbf{b}$~are:
%
\begin{flalign}
\label{vector_bias}
   \frac {\partial \mathbf{x}_{R}}{\partial \mathbf{b}}  &=  \Big(\mathbf{\mathbf{H(x)}}^T \mathbf{\Sigma} \mathbf{\mathbf{H(x)}} \Big)^{-1}\mathbf{\mathbf{H(x)}}^T \mathbf{\Sigma}.
\end{flalign}

\subsection{Loss Function}

To evaluate the accuracy of the restored solution obtained from our proposed algorithm, we need to define a quantitative measure, i.e., a \emph{loss function}, that compares the restored solution to the true solution of the OPF problem. There are several possible ways to define a loss function in this context, such as comparing the voltage magnitudes, phase angles, power injections, line flows, etc. from the restored solution to those from the true OPF solution.

Following typical approaches for training ML models, we formulate a loss function as the squared difference between the voltage magnitudes and angles from the restored solution $\mathbf{x}_{R}$ and the true solution $\mathbf{x}_{AC}$. To achieve this, we introduce new vectors $\mathbf{X}_{R}= [{\mathbf{x}_{R}^{(1)}}^T, {\mathbf{x}_{R}^{(2)}}^T, \ldots, {\mathbf{x}_{R}^{(S)}}^T]^T$ and $\mathbf{X}_{AC}=[{\mathbf{x}_{AC}^{(1)}}^T, {\mathbf{x}_{AC}^{(2)}}^T, \ldots, {\mathbf{x}_{AC}^{(S)}}^T]^T$, where $\mathbf{x}_{R}^{(i)}$ and $\mathbf{x}_{AC}^{(i)}$ denote the vectors of voltage magnitudes and angles at each bus for the restored and actual OPF solutions of the $i$-th sampled load scenario and $S$ represents the number of scenarios. Consequently, we define the loss function across samples as:
%

%
\begin{equation}
\label{eq:objective2_norm} F(\mathbf{\Sigma}, \mathbf{b})=\frac{1}{n} \lVert \mathbf{X}_{R}(\mathbf{\Sigma}, \mathbf{b})-\mathbf{X}_{AC} \rVert_2^2,
\end{equation}
where $\frac{1}{n}$ normalizes this function based on the system size. 

\subsection{\textcolor{black}{Adaptive Stochastic Gradient Descent (ASGD) Algorithm}}

\textcolor{black}{The optimal weight and bias parameters, $\mathbf{\Sigma}$ and $\mathbf{b}$, are obtained using the ASGD method described in Algorithm~\ref{alg:proposed-method}. 
After the offline execution of Algorithm~\ref{alg:proposed-method}, the resulting weights are applied online to restore AC power flow feasibility for a particular problem via Algorithm~\ref{alg:NR} (see~Fig.~\ref{fig:flowchart}.)}

\textcolor{black}{To compute optimal weight and bias parameters, Algorithm~\ref{alg:proposed-method} first creates a set of sampled load scenarios representing the range of conditions expected during real-time operations. Next, the algorithm solves (in parallel) both the actual and simplified OPF problems and saves the results in $\mathbf{x}_{AC}^{(i)}$ and $\mathbf{z}^{(i)}$, respectively, for each load scenario $i=1,\ldots,S$. Using the information from the simplified solutions and Algorithm~\ref{alg:NR}, the algorithm computes the restored solutions $\mathbf{x}_{R}^{(i)}$ for each load scenario $i=1,\ldots,S$. The algorithm then iteratively updates the weight and bias parameters based on the discrepancies between the actual and restored solutions along with their respective partial derivatives in order to minimize the loss function. The optimal weight and bias parameters, $\mathbf{\Sigma}^{opt}$ and $\mathbf{b}^{opt}$, are returned as outputs after reaching a maximum number of iterations or satisfying some other termination criteria (e.g., negligible changes from one iteration to the next).}

The ASDG algorithm uses the gradient of the loss function with respect to the weight parameters, denoted as $\mathbf{q}^{var}$:
%

%
\begin{equation}\label{eq:gradient}
\mathbf{q}^{var}= \frac{2}{n}\sum_{i=1}^{S}\left.\frac{ \partial \mathbf{x}_{R}}{\partial \mathbf{\Sigma}}\right|_{\mathbf{x}_R^{(i)}}\Big(\mathbf{x}_{R}^{(i)}(\mathbf{\Sigma}, \mathbf{b})-\mathbf{x}_{AC}^{(i)}\Big).
\end{equation}
There are many variants of gradient descent algorithms, such as batch gradient, momentum, AdaGrad, Adam, etc., each of which has their own advantages and disadvantages. 
We use the Adam algorithm since we empirically found it to perform best for this application~\cite{taheri2023acc}. The Adam algorithm is commonly used for training machine learning models and involves the following steps at each iteration~\cite{kingma2014adam}:
%
\begin{subequations}
\label{eq:adam}
\begin{align}
 \mathbf{m} &\leftarrow \beta_1 \mathbf{m} + (1-\beta_1)\mathbf{q}, \\
 \boldsymbol{\tau} &\leftarrow \beta_2 \boldsymbol{\tau}+ (1-\beta_2)(\mathbf{q})^2,\\
 \mathbf{\hat{m}} &\leftarrow \frac{\mathbf{m}}{1-\beta_1^{k}}, \\
 \boldsymbol{\Gamma} &\leftarrow \frac{\boldsymbol{\tau}}{1-\beta_2^{k}},\\
 \mathbf{\Sigma} &\leftarrow \mathbf{\Sigma} - \eta \frac{\mathbf{\hat{m}}}{\sqrt{\boldsymbol{\Gamma}} + \epsilon},
\end{align}
\end{subequations}
where $\mathbf{m}$ and $\boldsymbol{\tau}$ are the first and second moments of the gradients at iteration $k$, $\eta$ is a learning rate (step size), $\beta_1$ and $\beta_2$ are exponentially decaying hyperparameters for the first and second moments, and $\epsilon$ is a small constant.

In addition, the gradient of the objective function with respect to the bias parameters is represented by $\mathbf{q}^{bias}$:
%
\begin{equation}\label{eq:gradient_bias}
\mathbf{q}^{bias}=\frac{2}{n}\sum_{i=1}^{S}\left.\frac{ \partial \mathbf{x}_{R}^{(i)}}{\partial \mathbf{b}}\right|_{\mathbf{x}_R^{(i)}} \Big(\mathbf{x}_{R}^{(i)}(\mathbf{\Sigma}, \mathbf{b})-\mathbf{x}_{AC}^{(i)}\Big).  
\end{equation}
Using this gradient, one can find the optimal bias parameters $\mathbf{b}$ using the Adam algorithm in the same fashion as in (\ref{eq:adam}).

\begin{algorithm}[t]
\label{alg:proposed-method}
\smaller 

\DontPrintSemicolon
\caption{Computing Weight and Bias Parameters}

\KwInput{$\eta, \epsilon, \beta_1, \beta_2, \mathbf{m}, \boldsymbol{\tau}, \textit{batch size}, \textit{max\_iter}$: Adaptive stochastic gradient descent parameters \\
\hspace{1.1cm}$\mathbf{\Sigma}^{init}$: Initial weight parameters \\
\hspace{1.1cm}$\mathbf{b}^{init}$: Initial bias parameters
}

\KwOutput{$\mathbf{\Sigma}^{opt}$: Optimal weight parameters \\
\hspace{1.1cm}$\mathbf{b}^{opt}$: Optimal bias parameters
}

\KwData{OPF problem data}
Generate load scenarios $s_i$ for $i= 1,\ldots,S$ \\

Solve OPF problem~\eqref{eq:opf} for each scenario $s_i$, $i=1,\ldots,S$, and store the results in $\mathbf{x}_{AC}^{(i)}$ \\

Solve simplified OPF problems for each scenario $s_i$, $i=1,\ldots,S$, and store the results in $\mathbf{z}^{(i)}$ \\

$\mathbf{\Sigma} \leftarrow \mathbf{\Sigma}^{init}$,
$\mathbf{b} \leftarrow \mathbf{b}^{init}$\\

$\mathbf{m}^{var} \leftarrow \mathbf{m}$, 
$\mathbf{m}^{bias} \leftarrow \mathbf{m}$, $\boldsymbol{\tau}^{var} \leftarrow \boldsymbol{\tau}$, $\boldsymbol{\tau}^{bias} \leftarrow \boldsymbol{\tau}$,
$k \leftarrow 1$

\While{\begin{math} k \leq max\_iter \end{math}}
{
$\mathbf{X}_{R} \leftarrow [~]$, 
$\mathbf{X}_{AC} \leftarrow [~]$,
$\mathbf{q}^{var} \leftarrow [~]$,
$\mathbf{q}^{bias} \leftarrow [~]$

    \For{$i \in \text{random.sample}(\left\lbrace 1,\ldots,S\right\rbrace\!\text{), batch size)}$ \emph{\textbf{(in parallel)}}} 
	{               

         Run the restoration method in Algorithm~\ref{alg:NR} for scenario $s_i$ with simplified OPF solution $\mathbf{z}^{(i)}$, weights $\mathbf{\Sigma}$, and bias $\mathbf{b}$ to an accuracy of $\varepsilon$ and store the solution in $\mathbf{x}_{R}^{(i)}$ \\
    	        
         $\mathbf{q}^{var} \leftarrow \mathbf{q}^{var} + \frac{2}{n}\left.\frac{ \partial \mathbf{x}_{R}}{\partial \mathbf{\Sigma}}\right|_{\mathbf{x}_R^{(i)}} (\mathbf{x}_{R}^{(i)}-\mathbf{x}_{AC}^{(i)})$ \\
                
         $\mathbf{q}^{bias} \leftarrow \mathbf{q}^{bias} + \frac{2}{n}\left.\frac{ \partial \mathbf{x}_{R}}{\partial \mathbf{b}} \right|_{\mathbf{x}_R^{(i)}}(\mathbf{x}_{R}^{(i)}-\mathbf{x}_{AC}^{(i)})$ \\
    	
     $\mathbf{X}_{AC} \gets \text{append}(\mathbf{X}_{AC}, \mathbf{x}_{AC}^{(i)})$ \\
     
     $\mathbf{X}_{R} \gets \text{append}(\mathbf{X}_{R}, \mathbf{x}_{R}^{(i)})$ \\
    }
    %
    
        $\mathbf{m}^{var} \gets \beta_1 \mathbf{m}^{var} + (1-\beta_1)\mathbf{q}^{var}$\hspace*{5em}\tikzmark{right}\tikzmark{top_var}
          
        $\boldsymbol{\tau}^{var} \gets \beta_2 \boldsymbol{\tau}^{var} + (1-\beta_2)\mathbf{q}^{var}$
        
        $\mathbf{\hat{m}}^{var} \gets \frac{\mathbf{m}^{var}}{1-\beta_1^{k}}$%
        
        $\boldsymbol{\Gamma}^{var} \gets \frac{\boldsymbol{\tau}^{var}}{1-\beta_2^{k}}$
        
        $\mathbf{\Sigma} \gets \mathbf{\Sigma} - \eta \frac{\mathbf{\hat{m}}^{var}}{\sqrt{\boldsymbol{\Gamma}^{var} + \epsilon}}$\tikzmark{bottom_var}
        \AddNote{top_var}{bottom_var}{right}{Adam method\\for weights $\mathbf{\Sigma}$}

    $\mathbf{m}^{bias} \gets \beta_1 \mathbf{m}^{bias} + (1-\beta_1)\mathbf{q}^{bias}$\tikzmark{top_bias}
    
    $\boldsymbol{\tau}^{bias} \gets \beta_2 \boldsymbol{\tau}^{bias} + (1-\beta_2)(\mathbf{q}^{bias})^2$
    
    $\mathbf{\hat{m}}^{bias} \gets \frac{\mathbf{m}^{bias}}{1-\beta_1^{k}}$
    
    $\boldsymbol{\Gamma}^{bias} \gets \frac{\boldsymbol{\tau}^{bias}}{1-\beta_2^{k}}$
    
    $\mathbf{b} \gets \mathbf{b} - \eta \frac{\mathbf{\hat{m}}^{bias}}{\sqrt{\boldsymbol{\Gamma}^{bias} + \epsilon}}$\tikzmark{bottom_bias}
    \AddNote{top_bias}{bottom_bias}{right}{Adam method\\for biases $\mathbf{b}$}

    \begin{math}
    k\leftarrow k+1
    \end{math}

}

$\mathbf{\Sigma}^{opt} \leftarrow \mathbf{\Sigma}$, \quad 
$\mathbf{b}^{opt} \leftarrow \mathbf{b}$ 
\end{algorithm}


\section{Experimental Results and Discussion}
\label{sec:Results and Discussion}

This section evaluates the proposed algorithm's performance using numerical results from restoring AC power flow feasibility for solutions obtained from the SOCP~\cite{jabr2006radial}, QC~\cite{coffrin2015qc}, and SDP~\cite{lavaei2011zero} relaxations, the LPAC approximation~\cite{coffrin2014lpac}, and the ML-based OPF model from~\cite{klamkin2022active}.

\subsection{Experiment Setup}

We generated 10,000 scenarios (8,000 for training and 2,000 for testing) for each of the \texttt{PJM 5-bus}, \texttt{IEEE 14-bus}, \texttt{IEEE 57-bus}, \texttt{IEEE 118-bus},  \texttt{Illinois 200-Bus}~\cite{birchfield2016grid}, \texttt{IEEE 300-bus}, and \texttt{Pegase 1354-bus} systems~\cite{pglib}. These scenarios were created by multiplying the nominal load demands by a normally distributed random variable with zero mean and standard deviation of $10\%$. We set the convergence tolerance $\epsilon$ to $10^{-6}$ in Algorithm~\ref{alg:NR}. Solutions to the OPF problems and the relaxations and approximations were computed using \texttt{PowerModels.jl}~\cite{coffrin2018powermodels} with the solvers Ipopt~\cite{wachter2006implementation} and Mosek
 on a computing node of the Partnership for an Advanced Computing Environment (PACE) cluster at Georgia Tech.
 This computing node has a 24-core CPU and 32~GB of RAM. We also imported the ML results for available test cases from~\cite{klamkin2022active}. The restoration algorithm was implemented in Python 3.0 using a Jupyter Notebook. The 
solution times of the AC-OPF model for given test cases are reported in Table \ref{tab:acopf_solution_times} using \texttt{PowerModels.jl}.

\begin{table}[ht]
\caption{Average Execution time per scenario for AC-OPF}
\centering
\begin{tabular}{@{}lrrrrrrr@{}}
\toprule
Test Case     & \texttt{5} & \texttt{14} & \texttt{57} & \texttt{118} & \texttt{200} & \texttt{300} & \texttt{1354} \\ \midrule
Time (s) &  0.113  & 0.127   &  0.174  & 0.403   & 0.449   & 0.929 &  6.654  \\ \bottomrule
\end{tabular}
\label{tab:acopf_solution_times}
\end{table}

\subsection{Benchmarking Approach}
We consider three alternate restoration methods as comparisons to our proposed algorithm. The first method simply compares the voltage magnitudes and angles from the relaxed, approximated, or ML-based solution directly to the OPF solution. However, it is important to note that this method typically does not yield an AC power flow feasible point and is thus unsuitable for many practical applications. Additionally, this method is not applicable to the SDP and SOCP relaxations as they do not have variables corresponding to the voltage phase angles.
The second method, referred to as the ``benchmark'' method, solves the power flow problem obtained from fixing the voltage magnitudes at all generator buses and the active power injections at non-slack generator buses to the outputs of the simplified OPF problem as discussed in~\cite{venzke2020inexact} and used in a variety of papers such as \cite{venzke2020inexact, Li2022, BOBO2021106625, VANIN2020106699, pan2022deepopf, Zamzam2020}. The third method is the proposed restoration algorithm with the initial weight and bias parameters $\mathbf{\Sigma}_{ii} = 1$ and $\mathbf{b}_i = 0$, $i=1,\ldots,m$. The fourth method is the proposed restoration algorithm with weight and bias parameters computed using Algorithm~\ref{alg:proposed-method}.

\subsection{Performance Evaluation}
\begin{figure*}[!t]
\centering
\includegraphics[width=3.5in]{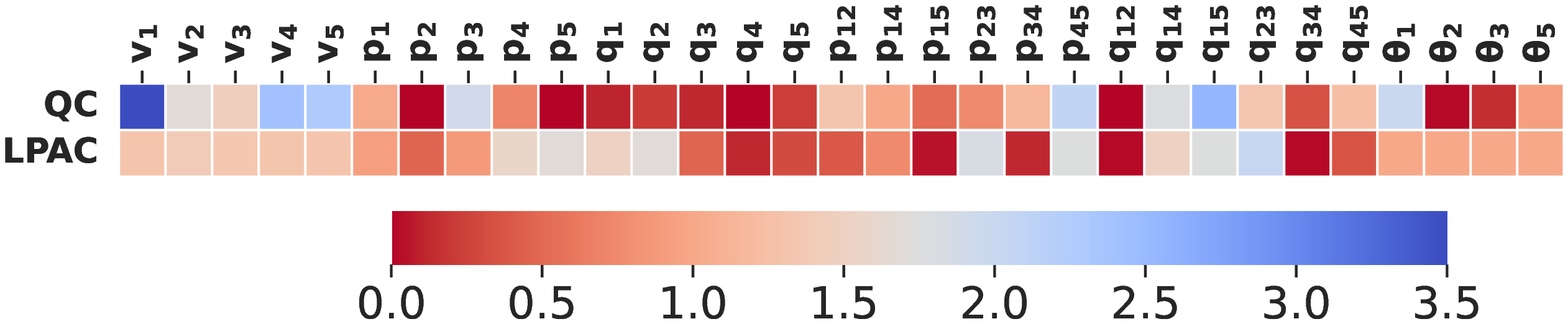}%
\label{fig_first_case}
\hfil
\includegraphics[width=3.5in]{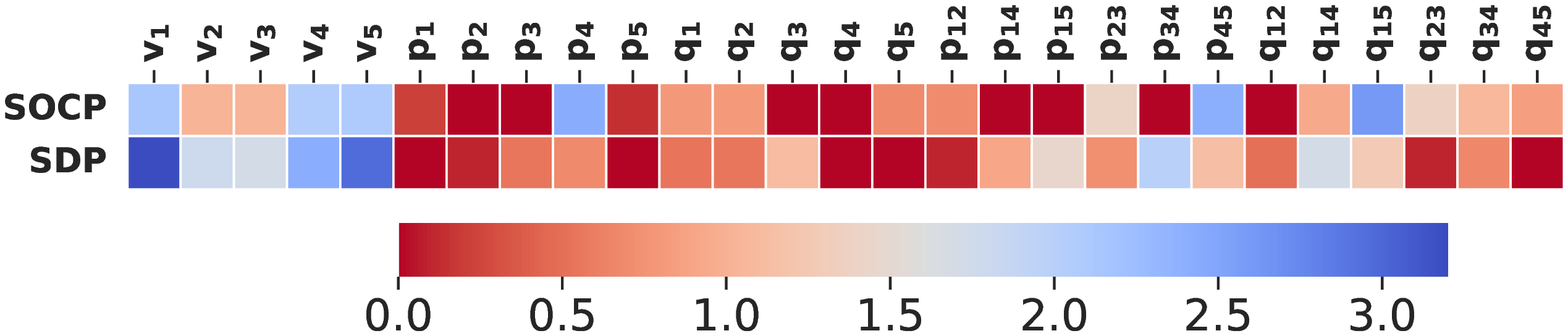}%
\label{fig_second_case}
\vspace{-0.8em}
\caption{Diagonal elements of the weight parameters $\Sigma$ for the 5-bus system as obtained using Algorithm~\ref{alg:proposed-method}. Higher values (in blue) indicate more trustworthy quantities for reconstructing AC feasible points. (a) QC and LPAC (left). (b) SOCP and SDP (right).}
\label{fig:weight_matrix}
\vspace{-1em}
\end{figure*}

\textcolor{black}{Fig.~\ref{fig:weight_matrix} shows the weight parameters (i.e., the diagonal elements of $\mathbf{\Sigma}$) obtained by applying Algorithm~\ref{alg:proposed-method} to the 5-bus system with the SOCP, QC, and SDP relaxations as well as the LPAC approximation. Observe that certain quantities receive significantly higher weights than others. For instance, in this test case, the algorithm allocates more weight to voltage magnitudes at buses 1 and 5, suggesting that these quantities are superior predictors of the actual OPF solutions. Larger weights imply that the algorithm considers these quantities more reliable when reconstructing the AC feasible points.}

\begin{figure*}[!t]
\centering
\subfloat[\smaller SOCP]{\includegraphics[width=1.7in, clip, trim=0 0 110 0 0]{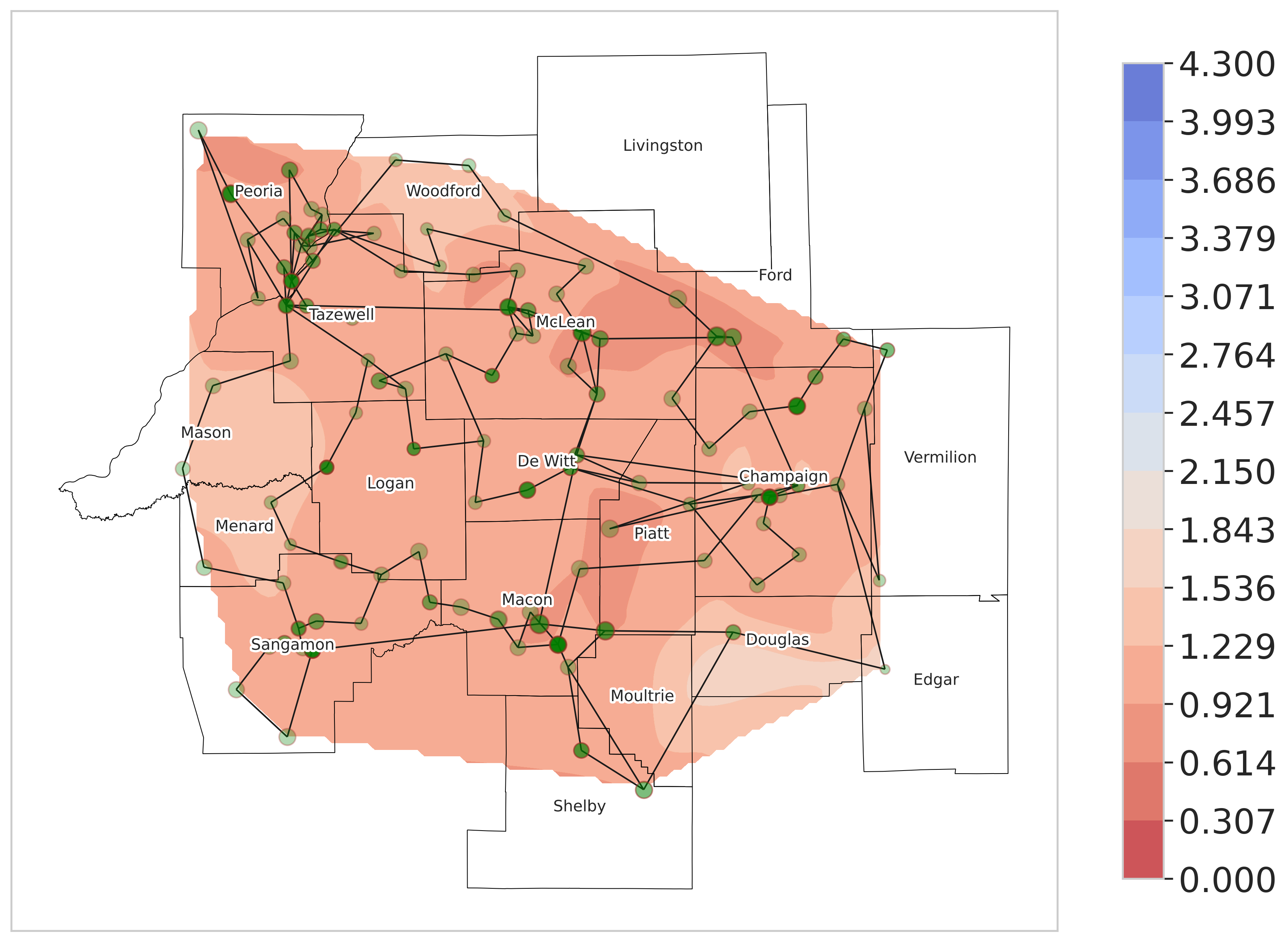}}%
\hfill
\subfloat[\smaller QC]{\includegraphics[width=1.7in, clip, trim=0 0 110 0]{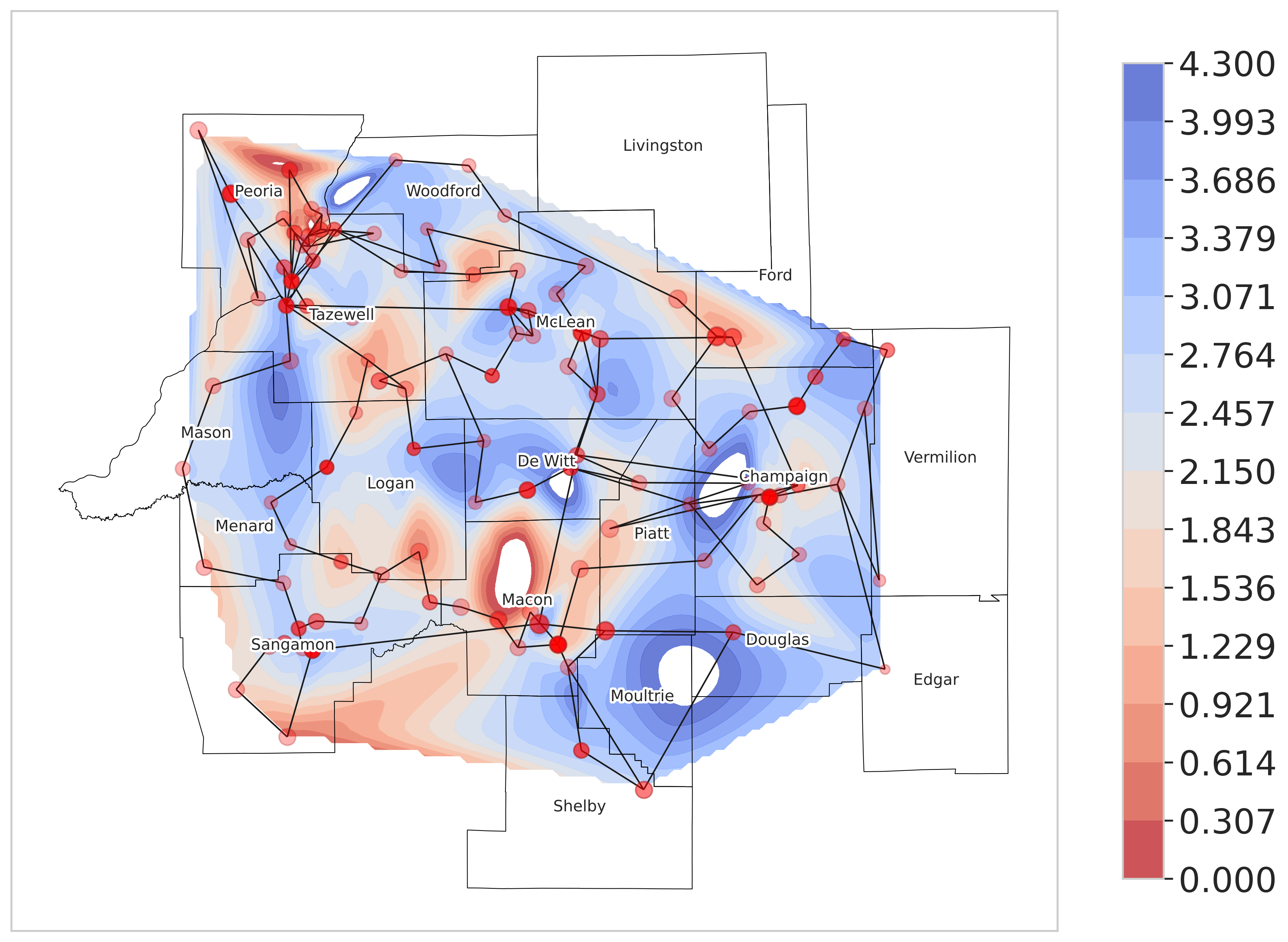}}%
\hfill
\subfloat[\smaller SDP]{\includegraphics[width=1.7in, clip, trim=0 0 110 0]{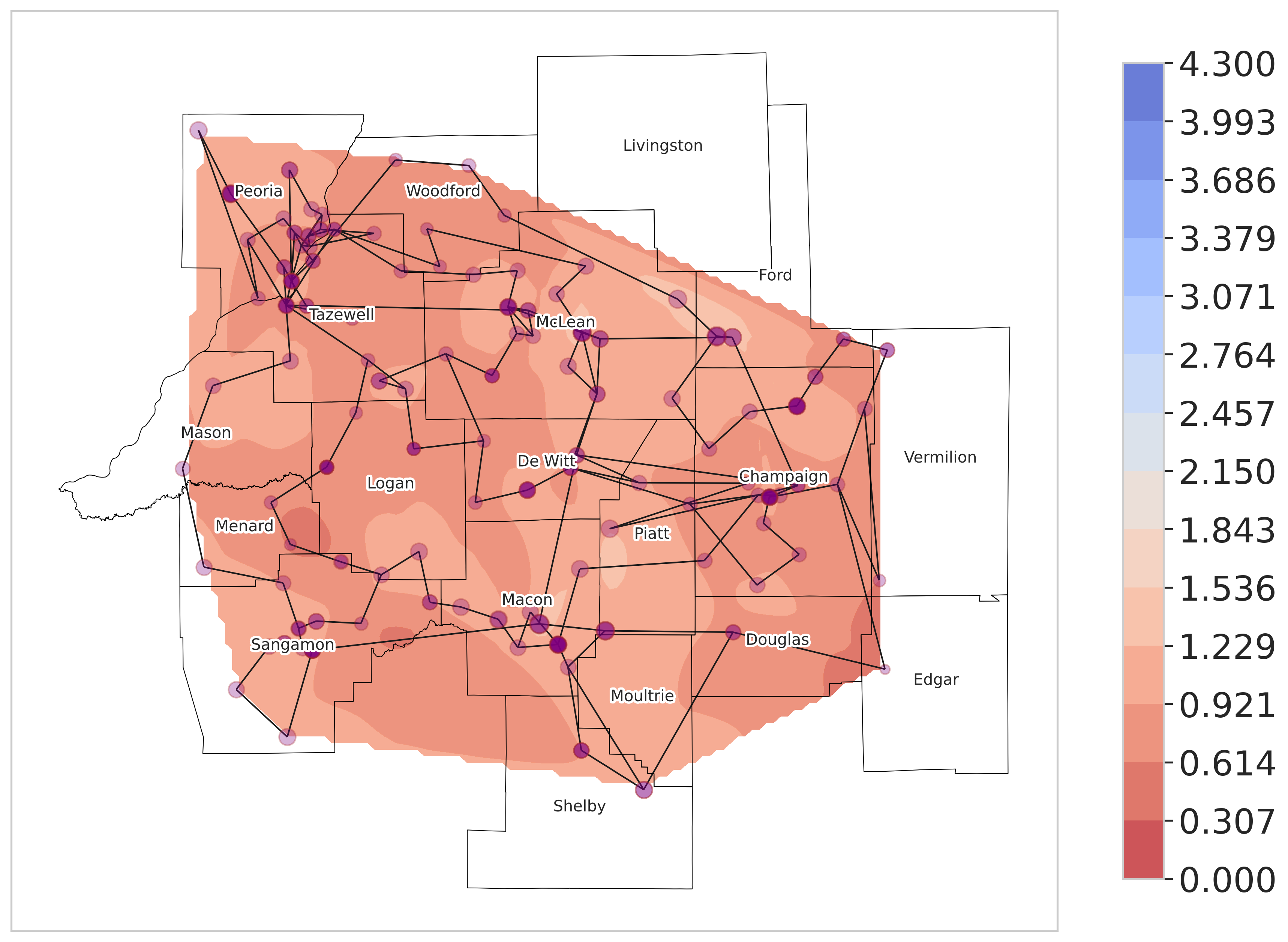}}%
\hfill
\subfloat[\smaller LPAC]{\includegraphics[width=1.96in, clip, trim=0 0 0 0]{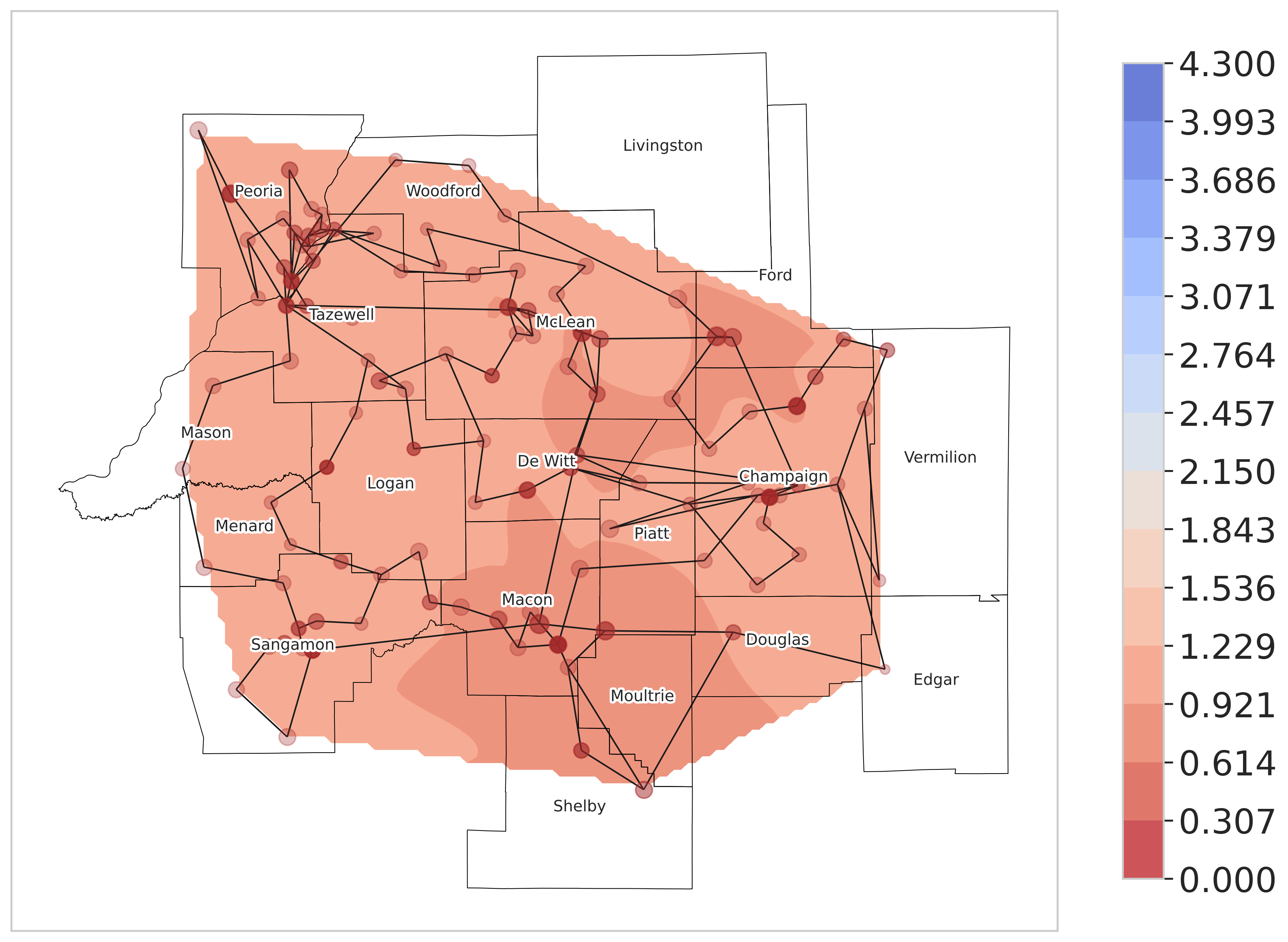}}%

\caption{Contour plot of weight parameters for voltage magnitudes in the Illinois 200-bus system for SOCP, QC, SDP, and LPAC.}
\label{fig:W_voltage_mag}
\vspace{-1em}
\end{figure*}

Moreover, Fig.~\ref{fig:W_voltage_mag} shows a geographic representation of the weight parameters $\mathbf{\Sigma}$ for the voltage magnitudes in the Illinois 200-bus system. The SOCP, QC, and SDP relaxations and the LPAC approximation each assign different weights to various parts of the system, with some clustering evident. Additionally, the QC relaxation has larger weights on the voltage magnitudes overall compared to the other OPF simplifications. These distinct weight assignments will be leveraged later in Section~\ref{subsec:combining} to combine multiple simplified OPF solutions for improved accuracy and performance, as our proposed algorithm can exploit the strengths of each method while compensating for individual inaccuracies.

\textcolor{black}{We evaluate the efficacy of the suggested restoration algorithm using the test dataset of 2,000 scenarios that were not used during the calculation of weight and bias parameters in Algorithm~\ref{alg:proposed-method}. Table~\ref{table:loss_function} presents the loss function values for each solution recovery method. As shown in Table~\ref{table:loss_function}, the proposed restoration algorithm successfully produces high-quality AC power flow feasible points from simplified OPF solutions. The loss functions resulting from the proposed algorithm are considerably smaller than those of other methods, including the benchmark approach. Furthermore, the application of optimized weight and bias parameters substantially enhances the performance of the loss function compared to using the initial weight and bias parameters $\mathbf{\Sigma}_{ii} = 1$ and $\mathbf{b}_i = 0$. Note that incorporating bias parameters into the updated algorithm further improves our initial findings presented in~\cite{taheri2023acc}.}

\begin{figure*}[!t]
\centering
\includegraphics[width=3.5in]{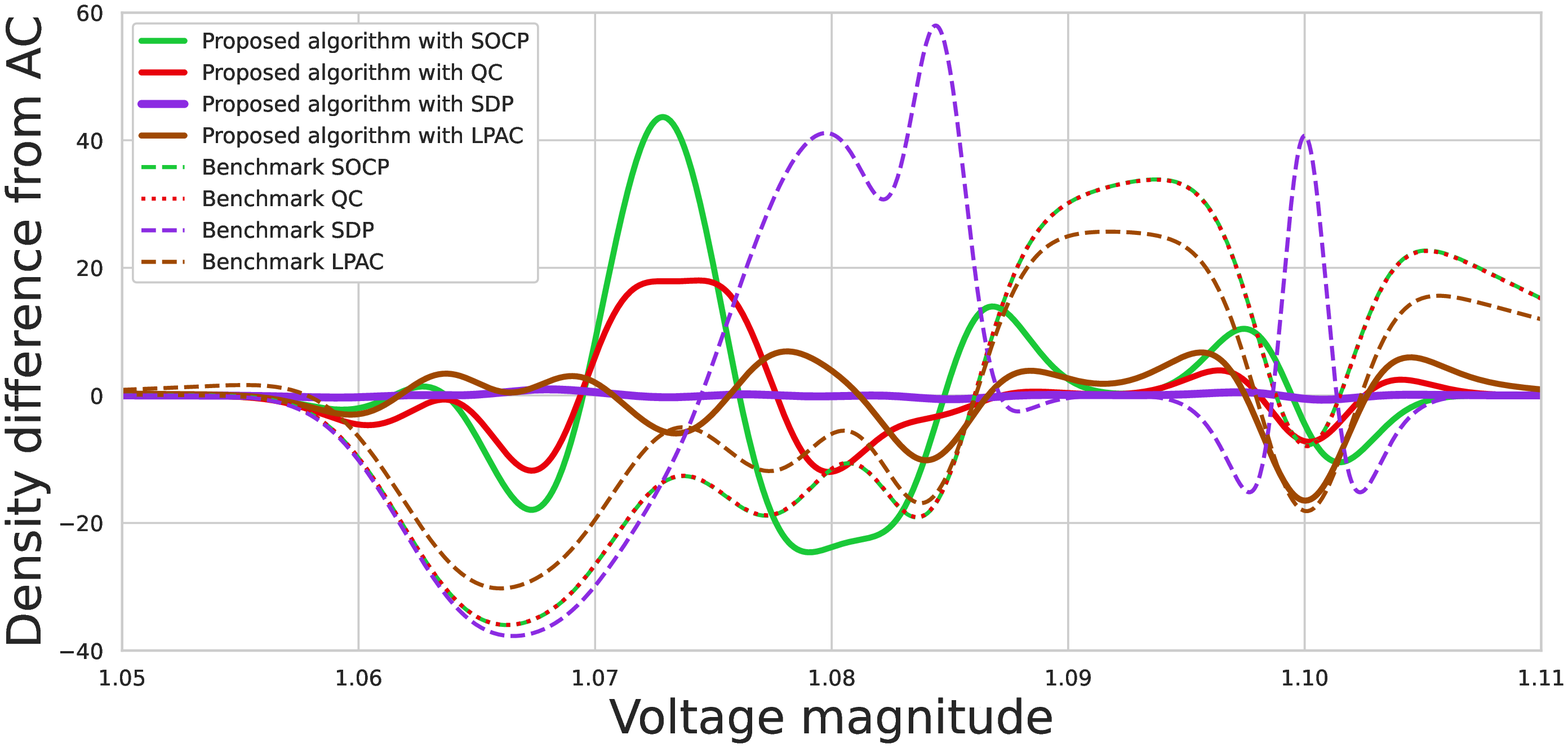}%
\label{fig:first}
\hfil
\includegraphics[width=3.5in]{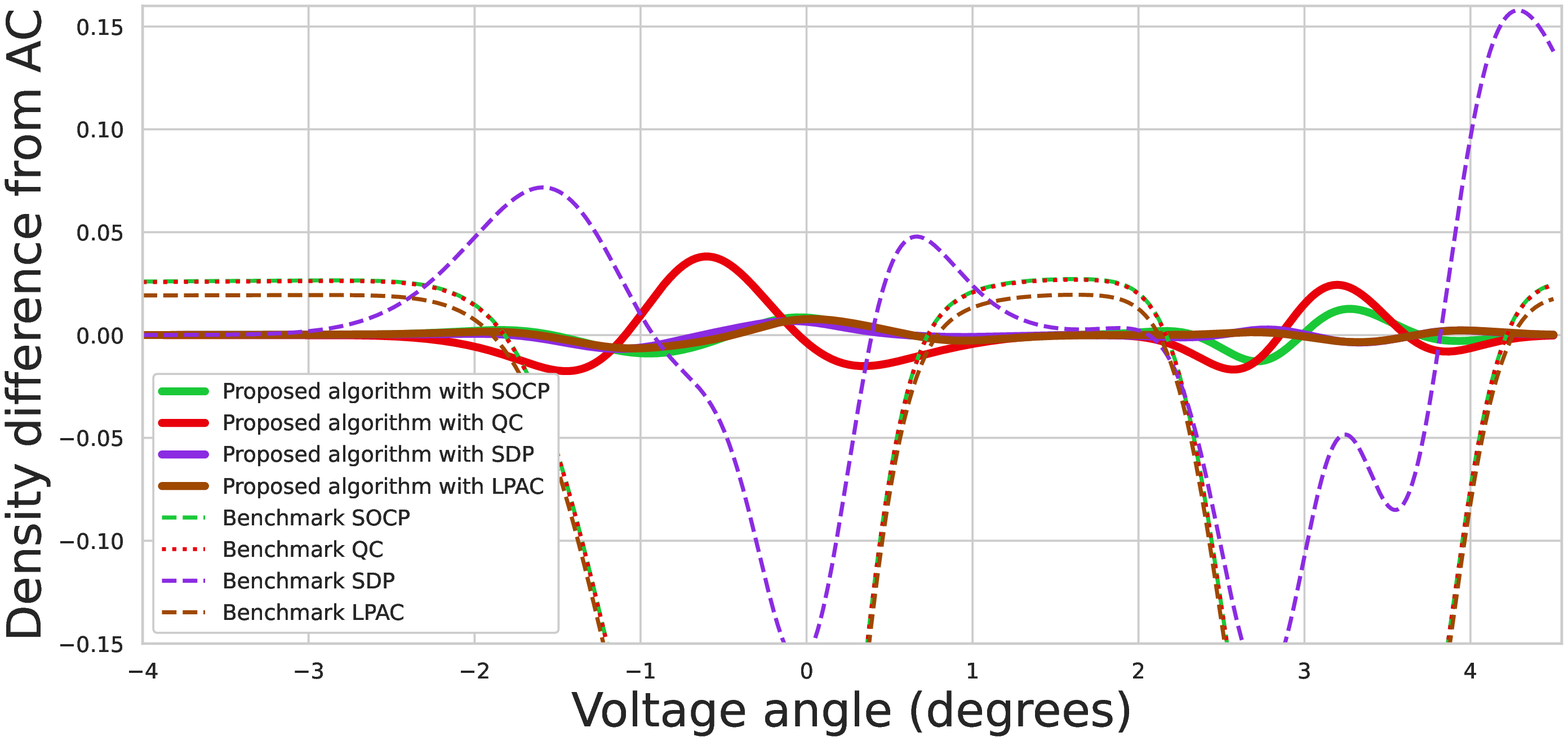}%
\label{fig:second}
\vspace{-0.8em}
\caption{Differences in the density distributions for the voltage magnitudes and angles for the restored points relative to the true OPF solutions. The vertical axis represents the difference in density from the true OPF solution, with positive values indicating higher density and negative values indicating lower density compared to true OPF solution for the 5-bus system over the test set of 2000 scenarios. (a) Voltage magnitudes. (b) Voltage angles (in degrees).}
\label{fig:SE_W_voltage_phasors_Combination}
\vspace{-1.2em}
\end{figure*}

Additionally, we compare the restoration methods by analyzing the difference in density distributions, which represent how frequently a particular voltage magnitude or angle value is observed in a restored solution relative to how often it is observed in the true OPF solution for the 5-bus system, considering all $2,000$ samples in the test dataset.
Fig.~\ref{fig:SE_W_voltage_phasors_Combination} demonstrates the performance of our proposed algorithm versus the benchmark, highlighting the superiority of the proposed algorithm when solving SDP relaxations of OPF problems. The figure has two subplots: (a) for voltage magnitudes and (b) for voltage angles. The vertical axes represent the difference in density relative to the true OPF solution, with positive values indicating higher density and negative values indicating lower density. Good performance is indicated by a line that is nearly horizontal at zero, which suggests that the restoration method accurately represents the true OPF solution across all voltage magnitudes and angles over $2,000$ test samples. This plot is useful for evaluating the overall performance of the restoration method, complementing the aggregate metrics in Table~\ref{table:loss_function} by assessing performance across voltage magnitudes and angles. We observe that the restored solutions to SDP relaxations (denoted by the solid purple line) outperform all other methods, including SOCP, QC, and LPAC, for this problem. Moreover, the proposed algorithm with SOCP, QC, SDP, and LPAC exhibits better performance than their respective benchmark counterparts, as indicated by the smaller differences relative to the true OPF solution.

\newcommand{\ra}[1]{\renewcommand{\arraystretch}{#1}}
\begin{table*}\centering
\caption{Loss Function Evaluated Using Test Dataset (2,000 scenarios) for Different Test Cases Using Various Simplified OPF Problems}
\ra{1.1}
\begin{tabular}{m{2.5cm}cccccccccc}
\toprule
\textbf{Test Case} & $\mathcal{|N|}$&$\mathcal{|L|}$&$\mathcal{|G|}$&$\mathcal{|E|}$&\textbf{Method} &\textbf{SOCP}& \textbf{QC}& \textbf{SDP} & \textbf{LPAC} &\textbf{ML} \cite{klamkin2022active}\\ \midrule 
 \multirow{4}{*}{\texttt {PJM 5-Bus}} 
& &&&&    Initial solution                    & ---   & 0.6709    & ---   &  0.4996     &--- \\
&\multirow{2}{*}{5}&\multirow{2}{*}{5}&\multirow{2}{*}{3}&\multirow{2}{*}{6}& Benchmark                           & 0.6077& 0.6069     & 0.1279  &0.4748  &---\\
&&&&&     SE with $\mathbf{\Sigma}^{init}$    &0.2355 & 0.2886  & 1.0840 & 0.2697      &---\\
&&&&&     SE with $\mathbf{\Sigma}^{opt}$     &\textbf{0.0055}& \textbf{0.0041}  & \textbf{0.0001} & \textbf{0.0033} &---\\

\cline{1-11}
\multirow{4}{*}{\texttt {IEEE 14-Bus}} 
&&&&&        Initial solution                      & ---            &  3.7926    & ---                 &  0.5655          &---\\
&\multirow{2}{*}{14}&\multirow{2}{*}{11}&\multirow{2}{*}{5}&\multirow{2}{*}{20}& Benchmark                             & 0.0010         &  0.0009    &\hphantom{00}\textbf{0.000003} & 0.1937  &---\\
&&&&&        SE with $\mathbf{\Sigma}^{init}$      & 0.2457         &  0.2284  & 0.2540  & 0.6110  &--- \\
&&&&&        SE with $\mathbf{\Sigma}^{opt}$       &\textbf{0.0002} &  \hphantom{1}\textbf{0.00008}  &\hphantom{1}0.00007 &\textbf{0.0012}   &--- \\

\cline{1-11}
\multirow{4}{*}{\texttt {IEEE 57-Bus}} 
&&&&&           Initial solution                     & ---    & 1.8558 & ---     & 0.5538    &---\\
&\multirow{2}{*}{57}&\multirow{2}{*}{42}&\multirow{2}{*}{7}&\multirow{2}{*}{80}&    Benchmark                            & 0.0566 & 0.0567 &  0.0463 & 0.7968    &--- \\
&&&&&           SE with $\mathbf{\Sigma}^{init}$     & 1.4155 & 1.3544 &  1.0713 & 2.4205    &---\\
&&&&&           SE with $\mathbf{\Sigma}^{opt}$      & \textbf{0.0099} & \textbf{0.0097} & \textbf{0.0091}  & \textbf{0.0214}  &---\\

\cline{1-11}
\multirow{4}{*}{\texttt {IEEE 118-Bus}} 
&&&&&           Initial solution                       & ---    &  6.1651 & ---     & 4.8066  &---\\
&\multirow{2}{*}{118}&\multirow{2}{*}{99}&\multirow{2}{*}{54}&\multirow{2}{*}{186}& Benchmark                              & 0.2056 & 0.2051  &  0.0113 & 5.0810  &---\\
&&&&&           SE with $\mathbf{\Sigma}^{init}$       & 7.3201 & 5.2822  & 6.8255  & 4.5119  &---\\
&&&&&           SE with $\mathbf{\Sigma}^{opt}$        & \textbf{0.0116}  &\textbf{0.0106}   & \textbf{0.0078}  & \textbf{0.0910}   &---\\

\cline{1-11}
\multirow{4}{*}{\texttt {Illinois 200-Bus}} 
&&&&&           Initial solution                       & ---    &  1.5836 & ---     & 10.8926  &---\\
&\multirow{2}{*}{200}&\multirow{2}{*}{108}&\multirow{2}{*}{49}&\multirow{2}{*}{245}&Benchmark                              & 0.1455 & 0.1492  &  0.1461 & 12.0743  &---\\
&&&&&           SE with $\mathbf{\Sigma}^{init}$       & 0.0024 &1.3533  &  0.0020  &10.9523  &---\\
&&&&&           SE with $\mathbf{\Sigma}^{opt}$         & \textbf{0.0001} &\textbf{0.0004}  & \textbf{0.0004}  & \textbf{0.0053}   &---\\

\cline{1-11}
\multirow{4}{*}{\texttt {IEEE 300-Bus}}
&&&&&            Initial solution                     & ---         &\hphantom{1}4.7658                  & ---    & 10.5747 & 0.5284\\
&\multirow{2}{*}{300}&\multirow{2}{*}{201}&\multirow{2}{*}{69}&\multirow{2}{*}{411}& Benchmark                            &10.5438      &\hphantom{1}8.8866               &37.7240 & 19.9149 &6.9820\\
&&&&&            SE with $\mathbf{\Sigma}^{init}$     &35.7829         &19.3754                &31.9334 & 10.2737 & 0.7901\\
&&&&&            SE with $\mathbf{\Sigma}^{opt}$      &\hphantom{2}\textbf{0.1891}     &\hphantom{1}\textbf{0.1602}     &\hphantom{1}\textbf{0.8809}  &\hphantom{1} \textbf{0.9242}&\textbf{0.1702}\\

\cline{1-11}
\multirow{4}{*}{\texttt{Pegase 1354-Bus}} 
&&&&&               Initial solution                     & ---            &10.7125                     & ---                         &10.3606  & 0.0568 \\
&\multirow{2}{*}{1354}&\multirow{2}{*}{673}&\multirow{2}{*}{260}&\multirow{2}{*}{1991}& Benchmark                            &3.7920          &\hphantom{1}3.7415          &58.4361                      &20.2537  & 0.1561 \\
&&&&&               SE with $\mathbf{\Sigma}^{init}$     &9.2313          &\hphantom{1}8.9304          &16.3797                      &22.3679  & 0.0502  \\
&&&&&               SE with $\mathbf{\Sigma}^{opt}$      &\textbf{0.3602} &\hphantom{1}\textbf{0.3442} &\hphantom{1}\textbf{0.9623}  &\hphantom{1}\textbf{1.9174}   &  \textbf{0.0291}   \\
\bottomrule
\end{tabular}
\begin{minipage}{16.125cm}
\vspace{0.25em}
\footnotesize{Note that the loss function values in this table correspond to the error of the restored solution with respect to the true OPF solution as defined in~\eqref{eq:objective2_norm}, \emph{not} the convergence tolerance. All results (except for the initial solution) satisfy AC power flow feasibility to a tolerance $\epsilon = 10^{-6}$.}
\end{minipage}
\label{table:loss_function}
\ifarxiv
\else
\vspace{-1em}
\fi
\end{table*}

\begin{figure*}[!t]
\centering
\includegraphics[width=3.5in]{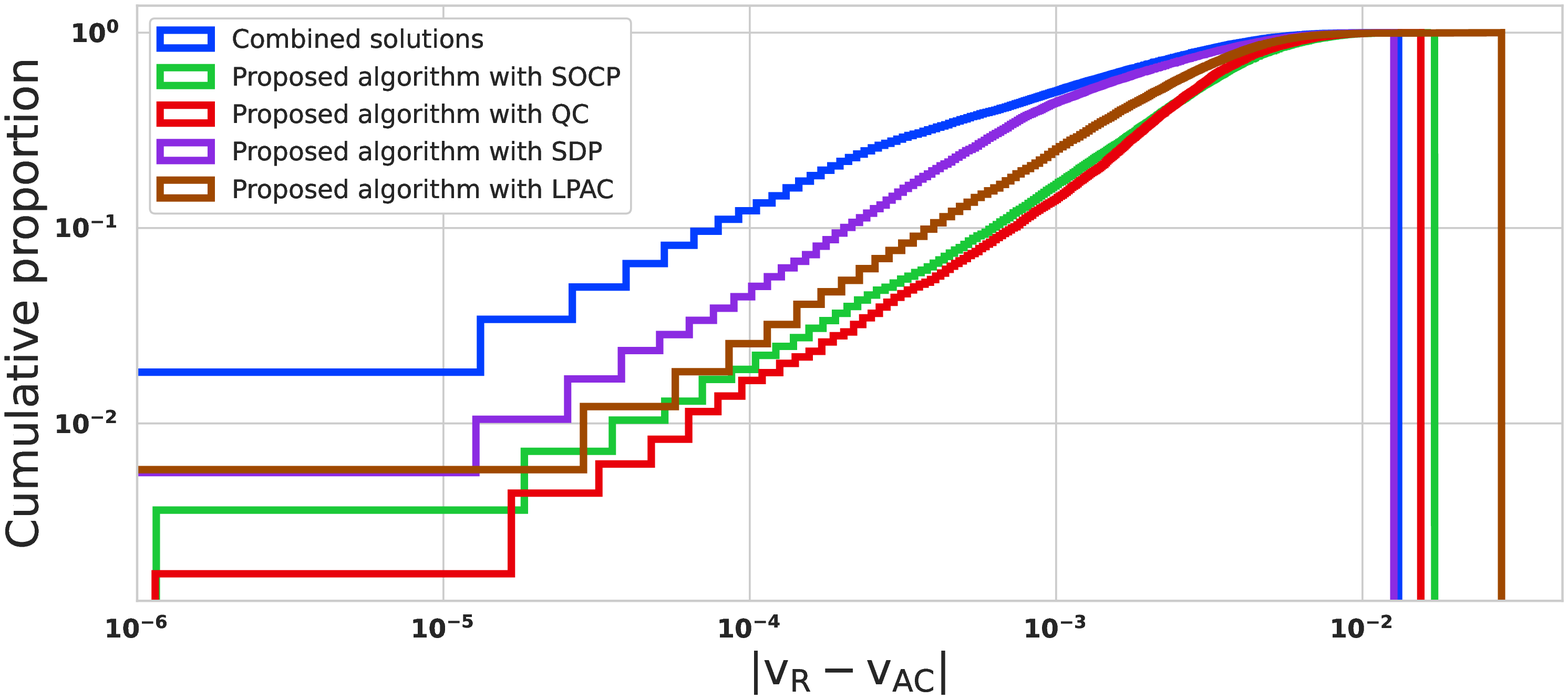}%
\label{fig:error_first}
\hfil
\includegraphics[width=3.5 in]{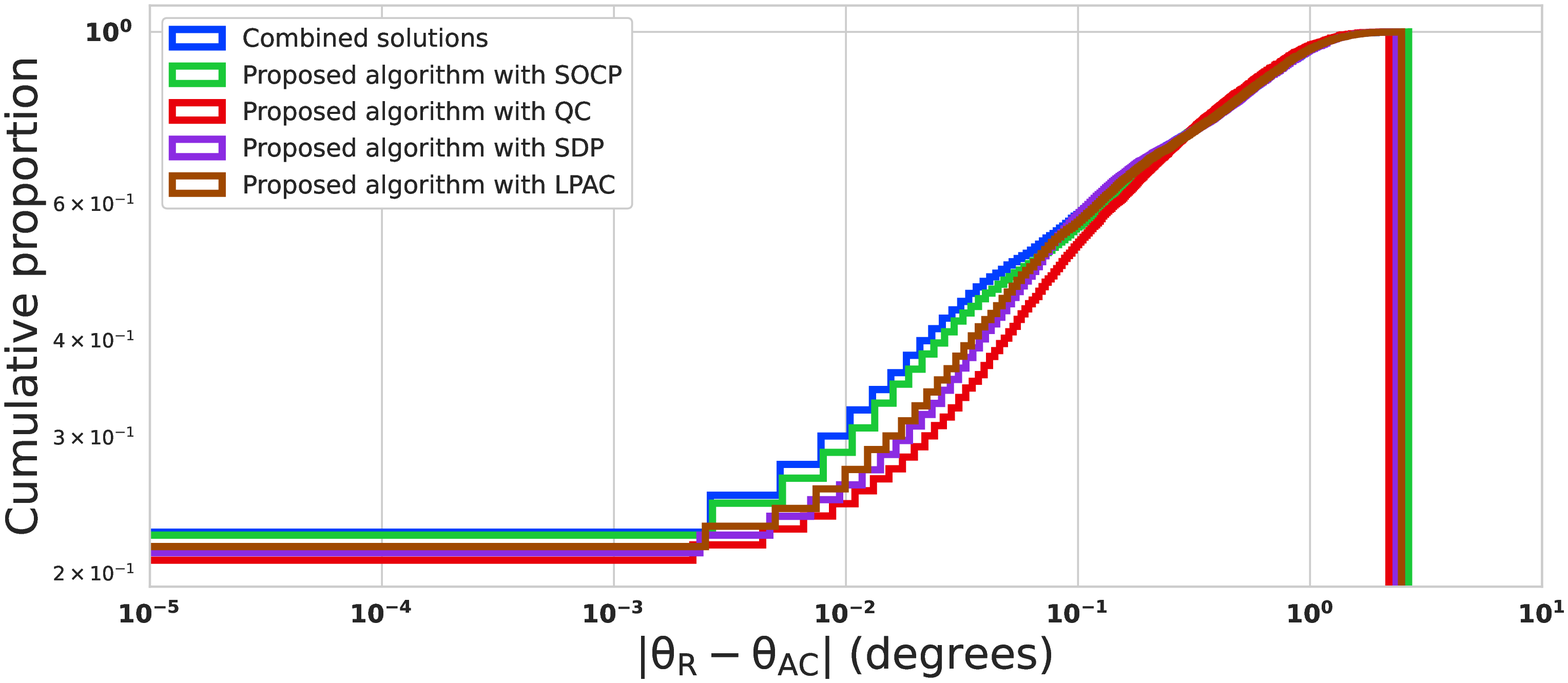}%
\label{fig:error_second}
\vspace{-0.8em}
\caption{Cumulative proportion of the absolute error in the voltage magnitudes and angles for various restoration methods for the 5-bus system over $2,000$ test scenarios. The vertical axis displays the cumulative proportion of the absolute error and the horizontal axis shows differing levels of the absolute error. Both the horizontal and vertical axes are on logarithmic scales. Each curve shows the cumulative proportion of errors up to a certain level, with higher curves thus indicating a larger proportion of smaller errors (i.e., better performance). (a) Voltage magnitudes. (b) Voltage angles (in degrees).}
\label{fig:error}
\vspace{-1.2em}
\end{figure*}

\subsection{Combining Solutions}\label{subsec:combining}
\textcolor{black}{The results presented above show that the restoration algorithm's outcomes vary with the choice of relaxation, approximation, or ML model, with none consistently dominating the others in every aspect. This suggests that there may be advantages in simultaneously considering \emph{multiple} simplified OPF solutions by combining their outputs using our proposed algorithm. 
With the flexibility to individually assign weights and biases for each quantity in a merged set of simplified OPF solutions, our proposed algorithm can naturally exploit the most accurate aspects of each simplified OPF solution while counterbalancing their individual inaccuracies.}
In other words, merging solutions from multiple simplified OPF problems supplies additional information for reconstructing even higher quality solutions via our proposed algorithm. For example, if we simultaneously use the results of the SOCP, QC, and SDP relaxations as well as the LPAC approximation, Algorithm~\ref{alg:proposed-method} will consider
{\footnotesize$ \mathbf{z}=[\mathbf{V}^T_{SOCP}~\mathbf{V}^T_{QC}~\mathbf{V}^T_{SDP}~\mathbf{V}^T_{LPAC}~\mathbf{P}^T_{SOCP}~\mathbf{P}^T_{QC}~\mathbf{P}^T_{SDP}~\mathbf{P}^T_{LPAC}~\mathbf{Q}^T_{SOCP}\\
~\mathbf{Q}^T_{QC}~\mathbf{Q}^T_{SDP}~\mathbf{Q}^T_{LPAC}~{\mathbf{P}^f}^T_{SOCP}~{\mathbf{P}^f}^T_{QC}~{\mathbf{P}^f}^T_{SDP}~{\mathbf{P}^f}^T_{LPAC}~{\mathbf{Q}^f}^T_{SOCP}\\
{\mathbf{Q}^f}^T_{QC}~{\mathbf{Q}^f}^T_{SDP}~{\mathbf{Q}^f}^T_{LPAC}~\boldsymbol{\theta}^T_{QC}~\boldsymbol{\theta}^T_{LPAC}]^T$}.
Algorithm~\ref{alg:proposed-method} automatically identifies the accuracy for each quantity in all simplified OPF solutions by suitably assigning the corresponding $\mathbf{\Sigma}_{ii}$ and $\mathbf{b}_i$ values to optimally utilize all available information, thus enabling recovery of high-quality solutions even when individual simplified OPF solutions might falter. For instance, the loss function for the 5-bus system with all simplified OPF solutions improves by an order of magnitude  relative to the restoration achievable with the best individual solution ($1 \times10^{-5}$ for the merged solutions versus $1 \times10^{-4}$ for the solution restored from the SDP solution alone).

Fig.~\ref{fig:error} compares the absolute errors in voltage magnitudes and angles, expressed as $\left|{\mathbf{X}_{AC} - {\mathbf{X}_{R}}}\right|$, for the 5-bus system over $2,000$ test samples. The horizontal axis denotes the absolute error and the vertical axis represents the cumulative proportion of errors less than or equal to the respective value on the horizontal axis. Ideal performance, i.e., a high proportion of small errors, would correspond to a curve in the upper-left corner.
With the steepest rise in Fig.~\ref{fig:error}, using multiple relaxations and approximations consistently leads to superior performance relative to using any single relaxation or approximation. 

Table~\ref{table:comput} gives online execution per-scenario run time with optimized weights and biases. The online execution is comparable to a power flow solve. For instance, \texttt{PowerModels.jl} performs the power flow calculations used by the benchmark restoration method in an average of 1.18 seconds for each scenario with the Pegase 1354-bus system, which is similar to the 1.02 to 1.21 seconds for our online execution.

%
%

\begin{table}\centering
\vspace*{-0.70em}
\caption{Average Execution Time per Scenario (seconds)}
\ra{1.0}
\begin{tabular}{m{2.5cm}ccccc}
\toprule

\textbf{Test Case} & \textbf{SOCP} &\textbf{QC}& \textbf{SDP} & \textbf{LPAC} \\ \midrule
\multirow{1}{*}{\texttt{PJM 5-Bus}} &
0.001 &0.002 &  0.001 &0.002\\

\hline
\multirow{1}{*}{\texttt{IEEE 14-Bus}} &
0.011 &0.008 & 0.013 & 0.009 \\

\hline
\multirow{1}{*}{\texttt{IEEE 57-Bus}} &
0.022 &0.019 & 0.023 & 0.021\\

\hline
\multirow{1}{*}{\texttt{IEEE 118-Bus}} &
 0.151 & 0.078 &0.158 & 0.081 \\

\hline
\multirow{1}{*}{\texttt{Illinois 200-Bus}} &
 0.230 &0.119 &0.243 &0.154 \\

\hline
\multirow{1}{*}{\texttt{IEEE 300-Bus}} &
0.310 &0.191 & 0.322 &0.217 \\

\hline
\multirow{1}{*}{\texttt{Pegase 1354-Bus}} &
1.132 &1.093 & 1.214 & 1.021 \\
\bottomrule
\vspace{-1em}
\end{tabular}
\label{table:comput}
\vspace{-2.5em}
\end{table}

\vspace{-0.4em}
\section{Conclusion}
\label{sec:Conclusion}

In this paper, we present a solution restoration algorithm that significantly improves AC power flow feasibility restoration from the solutions of simplified (relaxed, approximated, and ML-based) OPF problems. Our algorithm, which is based on state estimation techniques with the adjustment of weight and bias parameters through an ASGD method, can be several orders of magnitude more accurate than alternate methods. 

\textcolor{black}{In future research, we plan to utilize the insights gained from the trained weights and biases to improve the accuracy of power flow relaxations, approximations, and machine learning models. We also intend to incorporate the proposed restoration process into the training of machine learning models, thus closing the loop between training ML models and restoring AC feasible solutions.
Additionally, future research will aim to devise related self-supervised restoration techniques that do not depend on the availability of accurate OPF solutions. Instead, these techniques would instead compute the sensitivities of a solution quality metric with respect to the weight and bias parameters without requiring an optimal solution. The ability to restore high-quality AC power flow feasible operating points without the need for true OPF solutions would further increase the range of practical applications for the proposed algorithm.}

\section*{Acknowledgement}
The authors thank M.~Klamkin and P.~Van~Hentenryck for sharing the outputs of the ML models in~\cite{klamkin2022active}.

\bibliographystyle{IEEEtran}
\bibliography{refs}



\appendices

\section{Derivation of Sensitivities}\label{sec:sensitivity_derivation}

The sensitivities of the voltage phasors $\mathbf{x}_{R}$ obtained from the state estimation-inspired algorithm in relation to the weight matrix $\mathbf{\Sigma}$ are calculated using~\eqref{vector}, which is derived by taking the derivative of \eqref{eq:newton0} with respect to $\mathbf{\Sigma}$ as follows:
\begin{flalign}
   \label{eq:app1}
       &Y =(\mathbf{H}^{T} \mathbf{\Sigma} \mathbf{H})^{-1} \mathbf{H}^{T} \mathbf{\Sigma}(\mathrm{\mathbf{z}+\mathbf{b}}-\mathbf{h}),&
\end{flalign} 
 \begin{flalign}
  \label{eq:app2}
         dY&= \overbrace{d\Big(\left(\mathbf{H}^{T} \mathbf{\Sigma} \mathbf{H}\right)^{-1} \mathbf{H}^{T} \Big) \mathbf{\Sigma}\left(\mathrm{\mathbf{z}+\mathbf{b}}-\mathbf{h}\right)}^{I} \nonumber&\\
         &\qquad\mathrel{\phantom{=}}  + \underbrace{\left(\mathbf{H}^{T} \mathbf{\Sigma} \mathbf{H}\right)^{-1} \mathbf{H}^{T} d\Big(\mathbf{\Sigma}\left(\mathrm{\mathbf{z}+\mathbf{b}}-\mathbf{h}\right)\Big)}_{U},&
\end{flalign}
 \begin{flalign}
      \label{eq:app3}
        &U=\left(\mathbf{H}^{T} \mathbf{\Sigma} \mathbf{H}\right)^{-1} \mathbf{H}^{T} d\mathbf{\Sigma}\left(\mathrm{\mathbf{z}+\mathbf{b}}-\mathbf{h}\right),&
\end{flalign} 
 \begin{flalign}
    \label{eq:app4}
        &I=d\Big(\left(\mathbf{H}^{T} \mathbf{\Sigma} \mathbf{H}\right)^{-1}\Big) \mathbf{H}^{T}  \mathbf{\Sigma}\left(\mathrm{\mathbf{z}+\mathbf{b}}-\mathbf{h}\right),&
\end{flalign}
\begin{flalign}
   \label{eq:app5}
        I&=\Big(-\left(\mathbf{H}^{T} \mathbf{\Sigma} \mathbf{H}\right)^{-1} d\left (\mathbf{H}^{T} \mathbf{\Sigma} \mathbf{H}\right) \left(\mathbf{H}^{T} \mathbf{\Sigma} \mathbf{H}\right)^{-1}  \Big)\nonumber& \\
           &\qquad \qquad  \qquad \qquad\qquad\mathrel{\phantom{=}}  \times \Big(\mathbf{H}^{T}  \mathbf{\Sigma}\left(\mathrm{\mathbf{z}+\mathbf{b}}-\mathbf{h}\right)\Big),&
\end{flalign} 
\begin{flalign} 
   \label{eq:app6}
        I&=\Big(-\left(\mathbf{H}^{T} \mathbf{\Sigma} \mathbf{H}\right)^{-1} \mathbf{H}^{T} d\mathbf{\Sigma} \mathbf{H}  \left(\mathbf{H}^{T} \mathbf{\Sigma} \mathbf{H}\right)^{-1}  \Big) \nonumber& \\
            &\qquad \qquad  \qquad \qquad\qquad\mathrel{\phantom{=}}   \times \Big(\mathbf{H}^{T}  \mathbf{\Sigma}\left(\mathrm{\mathbf{z}+\mathbf{b}}-\mathbf{h}\right)\Big),&
\end{flalign} 
\begin{flalign}
   \label{eq:app7}
        dY=&-\left(\mathbf{H}^{T} \mathbf{\Sigma} \mathbf{H}\right)^{-1} \mathbf{H}^{T} d\mathbf{\Sigma} \mathbf{H} \left(\mathbf{H}^{T} \mathbf{\Sigma} \mathbf{H}\right)^{-1} \mathbf{H}^{T} \mathbf{\Sigma} \nonumber 
        (\mathrm{\mathbf{z}+\mathbf{b}}&\\
        & \qquad -\mathbf{h}) + \left(\mathbf{H}^{T} \mathbf{\Sigma} \mathbf{H}\right)^{-1} \mathbf{H}^{T} d\mathbf{\Sigma}\left(\mathrm{\mathbf{z}+\mathbf{b}}-\mathbf{h}\right),&
\end{flalign} 
\begin{flalign}
   \label{eq:app8}
        \textrm{vec}(dY)&=\textrm{vec} \Big [-\left(\mathbf{H}^{T} \mathbf{\Sigma} \mathbf{H}\right)^{-1} \mathbf{H}^{T} d\mathbf{\Sigma} \mathbf{H} \left(\mathbf{H}^{T} \mathbf{\Sigma} \mathbf{H}\right)^{-1}\nonumber&\\
           & \qquad\mathrel{\phantom{=}} \times \mathbf{H}^{T}  \mathbf{\Sigma}\left(\mathrm{\mathbf{z}+\mathbf{b}}-\mathbf{h}\right) \Big]+ \textrm{vec} \Big[ \left(\mathbf{H}^{T} \mathbf{\Sigma} \mathbf{H}\right)^{-1}\nonumber& \\
               & \qquad \qquad \quad \qquad\mathrel{\phantom{=}}  \times \mathbf{H}^{T}  d\mathbf{\Sigma} \left(\mathrm{\mathbf{z}+\mathbf{b}}-\mathbf{h}\right) \Big],&
\end{flalign} 
 \begin{flalign}
   \label{eq:app9}
      \textrm{vec}(dY)&= \Big [ - \Big(\mathbf{H} \left(\mathbf{H}^{T} \mathbf{\Sigma} \mathbf{H}\right)^{-1} \mathbf{H}^{T} \mathbf{\Sigma}  \left(\mathrm{\mathbf{z}+\mathbf{b}}-\mathbf{h}\right) \Big) ^{T} \nonumber& \\
      &\mathrel{\phantom{=}}\otimes \Big( \left(\mathbf{H}^{T} \mathbf{\Sigma} \mathbf{H}\right)^{-1} \mathbf{H}^{T} \Big) \Big] \textrm{vec} (d\mathbf{\Sigma})+\Big[  (\mathbf{z}+\mathbf{b}-\mathbf{h})^{T} \nonumber& \\
      &\mathrel{\phantom{=}}  \otimes \Big( \left(\mathbf{H}^{T} \mathbf{\Sigma} \mathbf{H}\right)^{-1} \mathbf{H}^{T} \Big)  \Big] \textrm{vec} (d\mathbf{\Sigma}),&
\end{flalign}
 \begin{flalign}
   \label{eq:app10}
     \textrm{vec}(dY)&= \Big [(\mathbf{z}+\mathbf{b}-\mathbf{h})^{T} \otimes \Big( \left(\mathbf{H}^{T} \mathbf{\Sigma} \mathbf{H}\right)^{-1} \mathbf{H}^{T} \Big)\nonumber&\\ 
        &\mathrel{\phantom{=}}- \Big(\mathbf{H} \left(\mathbf{H}^{T} \mathbf{\Sigma} \mathbf{H}\right)^{-1} \mathbf{H}^{T} \mathbf{\Sigma}\left(\mathrm{\mathbf{z}+\mathbf{b}}-\mathbf{h}\right) \Big) ^{T} \nonumber& \\
        &\mathrel{\phantom{=}}\otimes \Big( \left(\mathbf{H}^{T} \mathbf{\Sigma} \mathbf{H}\right)^{-1} \mathbf{H}^{T} \Big) \Big] \textrm{vec} (d\mathbf{\Sigma}), &
\end{flalign} 
 \begin{flalign}
  \label{eq:app11}
       \frac {\textrm{vec}(\partial Y)}{\textrm{vec}( \partial \mathbf{\Sigma})}  &= \Big ((\mathbf{z}+\mathbf{b}-\mathbf{h}) - (\mathbf{H}(\mathbf{H}^T \mathbf{\Sigma} \mathbf{H} )^{-1} \mathbf{H}^T \mathbf{\Sigma} \nonumber&\\
       &\quad \mathrel{\phantom{=}} \times (\mathbf{z}+\mathbf{b}-\mathbf{h})\Big)  \otimes \Big ((\mathbf{H}^T \mathbf{\Sigma} \mathbf{H})^{-1} \mathbf{H}^T \Big)^T.  &
\end{flalign}

\section{Exploiting the Diagonal Structure of $\Sigma$}\label{sec:diagonal}

We observe that the computation of~\eqref{eq:se_setup1} can be optimized by leveraging the diagonal structure of the $\mathbf{\Sigma}$ matrix. Instead of using a matrix, we can represent the weight terms as a vector $\boldsymbol{\sigma}$ with $\boldsymbol{\sigma}_{i} = \mathbf{\Sigma}_{ii}$, $i=1,\ldots,m$. This approach allows us to focus on computing sensitivities exclusively for the diagonal entries, leading to a more efficient calculation. We will next demonstrate this approach and its computational advantages.

First, we rewrite~\eqref{eq:se_setup1} as:
\begin{subequations}
\begin{align}
J(\mathbf{x}) &= (\mathbf{z}+\mathbf{b}-\mathbf{h(x)})^{T} \mathbf{\Sigma} (\mathbf{z}+\mathbf{b}-\mathbf{h(x)}), \\
\label{eq:ap_objective}
&= (\mathbf{z}+\mathbf{b}-\mathbf{h(x)})^{T} \boldsymbol{\sigma} \odot (\mathbf{z}+\mathbf{b}-\mathbf{h(x)}),
\end{align}
\end{subequations}
where $\odot$ denotes the Hadamard (element-wise) product. Following the same procedure as in \eqref{eq:se_g}--\eqref{eq:newton0}, we first compute the derivative of~\eqref{eq:ap_objective} with respect to $\mathbf{x}$ as follows:
\begin{equation}
\label{eq:app_g(x)}
\mathbf{g}(\mathbf{x})=\frac{\partial J(\mathbf{x})}{\partial \mathbf{x}} = -\Big(\boldsymbol{\sigma} \odot \mathbf{H(x)}\Big)^{T}(\mathbf{z}+\mathbf{b}-\mathbf{h(x)}),
\end{equation}
where $\mathbf{H(x)}=\frac{\partial \mathbf{h(x)}}{\partial \mathbf{x}}$ is the Jacobian matrix of the function $\mathbf{h(x)}$. The derivative of $\mathbf{g}(\mathbf{x})$ with respect to $\mathbf{x}$ is:

\begin{equation}
\mathbf{G(x)}=\frac{\partial \mathbf{g(x)}}{\partial \mathbf{x}} =
\Big(\frac{\partial \mathbf{h(x)}}{\partial \mathbf{x}}\Big)^{T} \boldsymbol{\sigma} \odot \mathbf{H(x)}.
\end{equation}

To solve $\mathbf{g(x)}=0$, we apply the Newton-Raphson method described in Algorithm~\ref{alg:NR} with modified $\mathbf{g(x)}$ and $\mathbf{G(x)}$ that performs, at the $k$-th iteration, the following steps:
\begin{subequations}
\begin{align}
\label{eq:app21}
\mathbf{x}^{k+1} &= \mathbf{x}^{k} - (\mathbf{G}(\mathbf{x}))^{-1}g(\mathbf{x}), \\
\label{eq:app22}
\Delta \mathbf{x} ^{k} &= \left(\mathbf{H}^{T} \boldsymbol{\sigma}  \odot \mathbf{\mathbf{H}}\right)^{-1} \Big(\boldsymbol{\sigma} \odot \mathbf{H}\Big)^{T}(\mathbf{z}+\mathbf{b}-\mathbf{h}), \\
\label{eq:app23}
\mathbf{x}^{k+1} &= \mathbf{x}^{k} + \Delta \mathbf{x} ^{k}.
\end{align}
\end{subequations}

Now, we can derive the sensitivities of the state vector $\mathbf{x}_{R}$ with respect to the $\boldsymbol{\sigma}$ in the same fashion as in Appendix~\ref{sec:sensitivity_derivation}:
\begin{flalign}\nonumber
   \frac {\partial \mathbf{x}_{R}}{ \partial \boldsymbol{\sigma}}  =&\Big((\mathbf{z}+\mathbf{b}-\mathbf{h}) - \big(\mathbf{\mathbf{H}}(\mathbf{\mathbf{H}}^T \boldsymbol{\sigma}\odot \mathbf{\mathbf{H}} )^{-1} 
    \mathbf{\mathbf{H}}^T \boldsymbol{\sigma} &\\ 
   & \quad \odot(\mathbf{z}+\mathbf{b}-\mathbf{h})\big)\Big) 
   \label{eq:app24}  \odot \Big((\mathbf{\mathbf{H}}^T \boldsymbol{\sigma}\odot \mathbf{\mathbf{H}})^{-1} \mathbf{H}^T \Big)^T.&
\end{flalign}

The expression~\eqref{eq:app24} gives the sensitivities of the restored point $\mathbf{x}_R$ with respect to the vector of weight parameters $\boldsymbol{\sigma}$. With length-$m$ vectors $\mathbf{z}$ and $\mathbf{h}(\mathbf{x})$ and an $m\times n$ matrix $\mathbf{H(x)}$, the sensitivities $\frac{ \partial \mathbf{x}_{R}}{\partial \boldsymbol{\sigma}}$ are represented by a $n \times m$ matrix. Accordingly, the size of the sensitivity matrix in~\eqref{vector}
reduces from $n \times m^2$ to $n \times m$; therefore, this approach results in a more efficient implementation than considering the sensitivities of all entries of the $\mathbf{\Sigma}$ matrix.

\end{document}